\documentclass[letterpaper,twocolumn,10pt]{article}
\usepackage{usenix}

\usepackage{tikz}
\usepackage{amsmath}

\usepackage{filecontents}

\usepackage[utf8]{inputenc}
\usepackage[T1]{fontenc}
\usepackage{subfiles}        
\usepackage{graphicx}
\usepackage{amsmath, amssymb}
\usepackage{booktabs}
\usepackage{enumitem}
\usepackage{hyperref}
\usepackage{xcolor}
\usepackage{microtype}
\usepackage{xspace}
\usepackage{longtable}
\usepackage{graphicx,xcolor,caption}

\newcommand{\reveree}{\textsc{Reveree}\xspace}
\newcommand{\dcipher}{\textsc{D-CIPHER}\xspace}

\usepackage{pifont}
\usepackage{subcaption}
\usepackage[normalem]{ulem}

\usepackage{graphicx,booktabs,array,multirow,xcolor,amsmath,caption}
\usepackage[most]{tcolorbox}

\definecolor{revgreen}{rgb}{0,0.5,0}

\usepackage{graphicx,booktabs,array,multirow,xcolor,amsmath,siunitx,caption}
\graphicspath{{rejudge_out/_results/}{rejudge_out/_crossset_figs/}{rejudge_out/rq3_name/}{rejudge_out/rq3_desc_divergent/}}

\hypersetup{
    colorlinks=true,
    linkcolor=blue,
    citecolor=blue,
    urlcolor=blue
}

\newif\ifshownotes
\shownotesfalse
\ifshownotes
  \newcommand{\gang}[1]{\textcolor{blue}{#1}}
  \newcommand{\hadjer}[1]{\textcolor{green}{#1}}
\else
  \newcommand{\gang}[1]{}
  \newcommand{\hadjer}[1]{}
\fi

\begin{document}

\title{Reveree: Diagnosing LLM Reverse-Engineering Agents}
\author{
{\rm Hadjer Benkraouda}\\
UIUC
\and
{\rm Hongyu Cai}\\
Purdue University
\and
{\rm Berkay Celik}\\
Purdue University
\and
{\rm Gang Wang}\\
UIUC
}
\maketitle

\begin{abstract}
Reverse engineering (RE) is critical to security tasks such as malware analysis and vulnerability discovery, and large language model (LLM) agents are increasingly able to perform it autonomously. Capture-the-flag (CTF) RE challenges have become the standard proxy for measuring this capability, but evaluation rests on a single criterion: whether the agent captures the flag. This solve rate reveals neither {\em where} in the RE process an agent fails nor {\em whether} a success reflects analysis of the binary or recall of a public solution. In this paper, we propose \reveree, a diagnostic framework that scores an LLM RE agent's trajectory at three tiers: solve rate, milestone progress through an eight-stage RE schema, and a behavioral profile of its actions. Comprehension stages are scored by an outcome-blinded LLM judge validated against a human expert; all other stages are verified deterministically. Using \reveree, we evaluate nine frontier models and four prompting strategies on 88 picoCTF and NYU-CTF challenges. We find that the base model dominates performance, whereas prompting strategy is a secondary, model-dependent effect. Surprisingly, larger, newer, or costlier models are not reliably stronger. We also find that failures concentrate at the comprehension stages of the RE process, and that extra budget, persistence, or reasoning effort rescues few of them, pointing to a competence limit rather than a resource limit. Regarding memorization, while models reproduce picoCTF flags from challenge descriptions alone, NYU-CTF shows minimal measurable recall, and most solves survive surface perturbation, indicating that genuine analysis coexists with memorization. We release \reveree to the community.

\gang{I feel ``security
tasks'' is too broad, maybe just say reverse-engineering tasks to begin with.}

\hadjer{indicating genuine reverse engineering rather than memorization.}
\gang{I feel this claim might be too strong (Lingming also does not believe it) --- we indeed find memorization on PicoCTF, and we should emphasize this.
Instead of claiming genuine reverse engineering, we may emphasize picCTF is suffer from memorization while NYU-CFY is less so (yet). This is still a useful conclusion. 
}
\end{abstract}

\section{Introduction}
\label{sec:intro}
Reverse engineering (RE), the recovery of a program's structure and behavior from its compiled form, is critical to security tasks such as malware analysis, vulnerability discovery, and firmware analysis~\cite{yong2021malware,malparse,sheng,vsp,costin2014firmware,eilam2011reversing, Benkraouda2025YouCJ}. The work is expensive and iterative: an analyst alternates between disassemblers, decompilers, debuggers, and scripts to recover what the code does and how to drive it~\cite{votipka2020observational,bryant2012sensemaking,ceccato2019understanding}. Large language models (LLMs) show promise on individual RE sub-tasks~\cite{llm4decompile,degpt,resym,gennm,symgen,deobfuscation}, and LLM \emph{agents}, which interleave reasoning with tool use, are increasingly able to carry out the process end to end~\cite{dcipher,enigma,pentestgpt,ctfagent,craken}. How well they do so is therefore a pressing question. The common way to answer it is to use capture-the-flag (CTF) RE challenges as a proxy~\cite{nyuctf,cybench,yang2023intercode,enigma,dcipher,turtayev,ctfjudge,vykopal2020ctf}: the agent is given a binary and must recover a hidden flag, a task that exercises real RE competencies, is machine-checkable, and can be perturbed without changing its difficulty.

Existing evaluations rest almost entirely on a single criterion, the \emph{solve rate}: the fraction of challenges whose flag is captured~\cite{nyuctf,cybench,yang2023intercode,enigma,dcipher,turtayev}. For reverse engineering, this criterion is necessary but insufficient. It \emph{floors} on hard challenges, where every model fails and diagnosis matters most. It cannot \emph{localize} failure: a run that recovered the binary's core algorithm and stalled one step short of a solver scores the same as one that never parsed the file, although the two call for different fixes. And it cannot separate \emph{reasoning from recall}: public challenges and their write-ups leak into pre-training, so a captured flag may be reproduced from memory rather than derived from the binary, a confound that grows as models train on more of the web content.

\noindent\textbf{Goal and Method.} In this paper, we develop \reveree, a diagnostic framework for evaluating LLM RE agents. Its key idea is to score the agent's \emph{trajectory} rather than its outcome alone, at three tiers: 1) the solve rate, 2) milestone progress through an eight-stage schema of the RE process. Stages that leave a mechanical trace are verified deterministically from tool calls; only comprehension stages are referred to an outcome-blinded LLM judge validated against a human expert, and
3) a behavioral profile that models the run as a Markov chain~\cite{markov} over behavioral states.
A memorization probe queries the bare model for a flag from the challenge description alone and re-runs the agent on renamed or paraphrased challenges, holding difficulty fixed. The harness is not our contribution: we build on \dcipher~\cite{dcipher} and hold it fixed, so that differences in outcome are attributable to the model or the prompting strategy rather than to the scaffold.

\noindent\textbf{Evaluation.} Using \reveree, we evaluate nine frontier models from the Claude and GPT-5 families, spanning size tiers and versions, under four prompting strategies (vanilla, ReAct~\cite{react}, Reflexion~\cite{reflexion}, and a hypothesis ledger) on 88 RE challenges from picoCTF~\cite{picoctf} and NYU-CTF~\cite{nyuctf}. \reveree exposes five patterns that solve rate hides. First, the base model dominates performance, whereas prompting strategy is model-dependent. Second, larger, newer, or costlier models are not reliably stronger. Third, failures are not random: on hard challenges, agents recover the structure of the binary and stall at the comprehension stages, and extra budget, persistence, or reasoning effort rescues few of them, pointing to a competence limit rather than a resource limit. Fourth, successful runs share a behavioral signature that failed runs lack, pivoting from analysis into building a solver rather than re-analyzing the binary, and it holds across every model and strategy. Finally, memorization is real but benchmark-dependent: models reproduce picoCTF flags from challenge descriptions alone, whereas NYU-CTF shows minimal measurable recall, and most solves survive surface perturbation, so genuine analysis coexists with memorization. These findings motivate RE-specific reasoning strategies aimed at the comprehension bottleneck (Section~\ref{sec:disc-prompting}).

\noindent\textbf{Contributions.} We make three contributions.
\begin{itemize}
\item We propose \reveree, an evaluation method beyond solve rate, that scores an agent's trajectory at three tiers and pairs them with a memorization probe. We open-source our tool to the community.
\item We perform a controlled evaluation of 
nine frontier models and four prompting strategies on 88 RE challenges and identify a success-versus-failure signature that replicates across all nine models.
\item We carryout a memorization evaluation to separate reasoning from recall with difficulty-preserving perturbations, and show that memorization is benchmark-dependent.
\end{itemize}
\gang{
After reading this sentence ``RE,  one of the CTF categories'', reviwers might have a simple question: why do you focus on RE among all these categories? Why not focus on all? Is RE the hardest? We should find a way to motivate ``why RE.'' A possible logic is to start with RE instead of starting with CTF: RE is important and LLM agent shows promise. We want to eval RE capability of LLM agent. The common way is to use CFT as a proxy. 
}

\gang{I think we should beef up this ``findings'' paragraph. If we can extract 3-4 (or even 5) super interesting findings, we can clearly list them with ``First'', ``Second'', ... ``Finally''. So readers can really understand the interesting/jucy part of the paper. Among these findings, we should discuss the memorization too, check my comment in the abstract. }

\gang{I am not sure if we can make the contribution items easier to understand. Maybe try to make the bold text a bit shorter. 
A new eval method beyond solve rate, Controlled model evaluation, Quantitative failure analysis, Memorization evaluation.  
}

\gang{is ``holding the harness fixed'' an advantage?'' People may argue this is a limitation since we did not test other harness? Maybe we should emphasize the ``Controlled'' eval? 
}

\gang{mention we share our code?}

\section{Background and Motivation}
\label{sec:background}
\label{sec:bg-re}
\label{sec:bg-ctf}
\label{sec:bg-agents}

\noindent{\bf Reverse engineering (RE).}
Reverse engineering is the recovery of a program's structure and behavior from
its compiled form, without access to source code.  It underpins a range of security tasks, including vulnerability discovery~\cite{sheng,vsp}, malware analysis~\cite{yong2021malware,malparse}, and firmware
analysis~\cite{costin2014firmware}; in many cases the only
available artifact to assess is a binary~\cite{chikofsky1990taxonomy, eilam2011reversing}. 
\hadjer{done}\gang{cite more papers for this statement.}
The work is inherently multi-step and iterative. Empirical studies of professional reverse engineers describe this trajectory: An analyst first \emph{triages} the binary to establish its format, architecture, and rough structure; \emph{maps its surface} through strings, imports, etc.,; \emph{recovers control flow} to be able to locate the functions that matter, i.e., \emph{recognizes the core algorithm} a target software implements. Next, the analyst \emph{extracts the constraints}. These are the conditions that a correct input must satisfy, e.g., a required byte at each position or the input/flag length that produce the desired behavior. Finally, they \emph{construct and run a solver} that produces the answer~\cite{votipka,bryant2012sensemaking,ceccato2019understanding}. 
\hadjer{done}\gang{It might be helpful to explain what a ``constraint'' means, maybe even with an example? }
\hadjer{done}\gang{We don't need to present two different ways for the same thing. Stick to one version (e.g., the version we used in the paper) and you can still cite these papers.}

Two features of this process
make it a stringent test of an autonomous agent. First, it relies on \emph{tool-use}:
progress depends on calling disassemblers, decompilers, and debuggers and
interpreting their output. Second, it is \emph{hypothesis-driven}, requiring the
analyst to form, test, and revise conjectures about what the code does rather than
to read an answer off the page. These sub-tasks are increasingly automated:
classical binary-analysis frameworks tackle them with symbolic and dynamic
techniques~\cite{angr,mayhem}, and a growing line of work applies LLMs to
decompilation, decompiler refinement, and variable- and function-symbol
recovery~\cite{llm4decompile,degpt,resym,gennm,symgen,deobfuscation,nova,bint5},
which sharpens the question this paper asks: whether an agent genuinely
comprehends a binary or merely pattern-matches its surface.
\hadjer{done}\gang{this is a good place for this paragraph I think}
 
\noindent{\bf CTFs as a proxy.}
\hadjer{done} \gang{We need to cite (many) related works that use CTF challenges as a proxy for LLM evaluation.}
Capture-the-flag (CTF) reverse-engineering challenges distill this process into a
controlled, verifiable task, and have become a standard proxy for evaluating LLM agents on offensive-security and reverse-engineering
tasks~\cite{nyuctf,cybench,yang2023intercode,enigma,dcipher,pentestgpt,turtayev,gioacchini2024autopenbench,muzsai2024hacksynth,autoattacker,craken,ctfjudge,ji2025ctf,Zhuo2025TrainingLM,anurin20243cb}. The agent is given a binary and a short description
and must recover a hidden \emph{flag}, a secret string typically requiring
understanding the binary's logic, such as a key-validation routine, a custom
encoding, or a layer of obfuscation. Three properties make CTF reverse
engineering a useful proxy for understanding real-world reverse engineering
activities, and account for its wide adoption as an evaluation
setting~\cite{vykopal2020ctf}. First, they exercise the same core competencies as real RE: static and dynamic
analysis, algorithm recognition, and constraint solving, across a wide range of difficulty. Second, its ground truth is unambiguous and machine-checkable: a
submitted flag either validates against the challenge oracle or does not, which
makes automated, reproducible evaluation possible without human grading. 
\hadjer{done}\gang{be careful with the term {\em at scale}. Is CTF a scalable way for evaluation? Especially given you mention the ``scale of real firmware '' below as a limitation.}
Third, challenges are
\emph{perturbable}: the binary can be modified while its analytic difficulty is held fixed, which is what makes a contamination-controlled experiment feasible. 
CTF challenges have some limitations; they omit the scale of real firmware 
and the open-ended goals of a genuine investigation.
\hadjer{done:removed}\gang{is it always true for the anti-analysis part?}
One property is double-edged and is central to our study: many challenges, and detailed write-ups of their solutions, are published online, so strong performance on a documented challenge may reflect recall of its solution rather than analysis of its binary.
 
\noindent{\bf LLM Agents.}
An LLM agent attempts an RE challenge by interleaving reasoning with tool use inside a
control loop. Given the binary, it may issue a tool call to, for example, disassemble a
function, decompile it, run it under a debugger, or execute a script. Next, it
observes the result, updates its understanding, and repeats until it submits a
flag or exhausts a budget of turns or tokens. A \emph{harness/framework} supplies
the scaffold around this loop. Single-agent harnesses give one model a curated
tool interface: EnIGMA~\cite{enigma}, built on SWE-agent~\cite{sweagent},
contributes interactive tools for debugging and remote interaction. 
Multi-agent harnesses decompose the task across roles: \dcipher~\cite{dcipher}, which we adopt, pairs a planner that sets
strategy with executors that issue the concrete tool calls. Beyond CTF challenges, agents are also used to automate penetration testing and broader security tasks~\cite{pentestgpt,autoattacker,ctfagent}. Orthogonal to the harness, the
agent's reasoning loop is shaped by general-purpose prompting strategies such as
ReAct~\cite{react}, which interleaves reasoning traces with actions, and
Reflexion~\cite{reflexion}, which adds self-reflection across attempts, among
others. 
Prompting strategies are typically
presented as model- and domain-agnostic. Across this landscape, agents are
evaluated on CTF benchmarks such as NYU-CTF~\cite{nyuctf} and almost always
by a single number: the solve rate~\cite{turtayev}.
\hadjer{done}\gang{We should explicitly mention the metric name, solve rate. }
 
\noindent{\bf Our Motivation.}
The bulk of the work on language-model agents for capture-the-flag and reverse
engineering is evaluated almost entirely by \emph{solve rate}: did the agent
submit the correct flag? For reverse engineering this single number is lossy
and insufficient, for three reasons that compound.
First, \textit{it floors on hard challenges.} A corpus of genuinely difficult binaries
drives solve rate toward zero for every model, collapsing the
distinctions between models, between strategies, between near-misses and
non-starters that an evaluation exists to draw. The harder and more interesting
the challenge, the less the solve rate metric reveals. 
Second, \textit{it cannot locate the cause of failures. }
Reverse engineering is a sequence of distinct sub-abilities, and an agent can fail at any of them; yet a run that recovered a function's algorithm and stalled at constraint extraction earns the same ``zero'' as a run that never parsed the binary, even though the two runs call for different fixes.
Third, \textit{it cannot separate reasoning from recall.} Because challenges and their
write-ups leak into pre-training, a captured flag may be reproduced from memory
rather than derived from the binary. This confound strengthens as models train on ever more of the public web, so an evaluation that ignores it becomes less trustworthy over time.
 
These gaps motivate a different kind of evaluation:
an instrument that (i) measures partial progress through the RE
process, so that hard challenges still discriminate; (ii) localizes where and how
an agent fails, so that failures are actionable rather than merely counted; and
(iii) separates genuine analysis from solution recall/memorization, so that a reported success can be trusted.
The remainder of this paper develops such an instrument, \reveree, and applies it.

\providecommand{\circled}[1]{\textcircled{\footnotesize#1}}

\begin{figure}[t]
\centering
\includegraphics[width=0.48\textwidth]{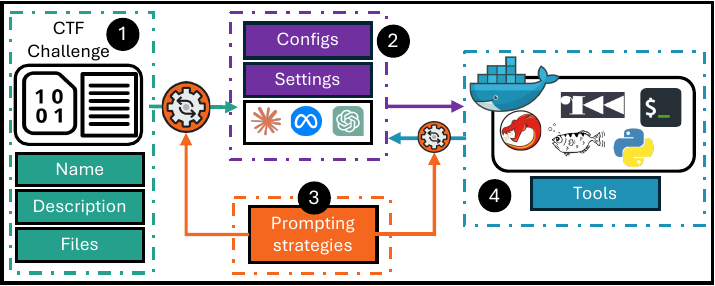}
\caption{The evaluation pipeline. A reverse-engineering challenge~\ding{182}
(name, description, files) is paired with a model configuration and
settings~\ding{183} and a prompting strategy~\ding{184}; the configured
agent runs in a sandboxed environment with a fixed reverse-engineering
toolset~\ding{185}, producing a trajectory that \reveree scores. The
 \emph{challenge features}, the \emph{model} and its \emph{configurations and settings}, and \emph{prompting strategies}
(\ding{182},~\ding{183},~\ding{184}) are the varied factors; the
execution environment and toolset (\ding{185}) are held fixed
across all runs.}
\label{fig:pipeline}
\end{figure}

\section{Evaluation Pipeline Overview}
\label{sec:pipeline}

Our study measures \emph{how} a language-model agent
attempts a reverse-engineering challenge by varying a set of factors. Figure~\ref{fig:pipeline} shows the pipeline that turns one factor setting into a scored run. A challenge~\ding{182}, comprising its name,
description, and binary files, is paired with a configuration~\ding{183} that
fixes the model and its hyperparameters and with a prompting
strategy~\ding{184} that determines the agent's reasoning loop. The configured
agent runs inside a sandboxed environment~\ding{185} equipped with a standard
reverse-engineering toolchain~\cite{ghidra, idapro, r2}, issuing tool calls and observing their output
until it submits a flag or exhausts its budget. The resulting trajectory is
scored by \reveree (Section~\ref{sec:reveree}); that scored trajectory is the pipeline's only output and the subject of measurement in this paper.

\subsection{Design Goals}
\label{sec:pipeline-goals}

We design the pipeline around three goals.

\begin{enumerate}
\item \textbf{Controlled variation.} To attribute a difference in
  behavior to a single factor (a model, a prompting strategy, a
  hyperparameter), every other part of the pipeline must be held fixed. The execution environment and the toolset are therefore
  identical across runs, so that a difference in outcome reflects the factor under study rather than a change in apparatus.
\item \textbf{Framework-agnostic.} The contribution is a methodology, not a
  harness/framework. We instantiate the pipeline on one agentic harness (D-CIPHER), but the parts
  that constitute the method do not depend on that harness. Only a simple adapter is needed to convert framework outputs into \reveree suitable inputs
  (Section~\ref{sec:reveree-inputs}). Any harness that records its tool calls and
  its outcome can be substituted without changing what is measured.
\item \textbf{A realistic agentic setting.} Agents act in a live environment with
  the disassemblers, decompilers, debuggers, and scripting tools a human reverse
  engineer would use, rather than answering questions about a static binary
  dump. What we measure is the agent's actual tool use, not a proxy for it.

\hadjer{[about 4th point]}
\gang{Need to be careful when explaining this claim. We need to clearly distinguish and clarify that this is about the reproducibility of \reveree's scoring, not the reproducibility of the way in which a model/agent solves a challenge. If I understand it correctly, a model may solve a challenge differently when running the same challenge multiple times?   
}\hadjer{you are right it is vague, removing for now - might bring back later}
\end{enumerate}

\subsection{The Agentic Framework}
\label{sec:pipeline-harness}

The agent is driven by D-CIPHER~\cite{dcipher}, a multi-agent CTF harness/framework in
which a planner directs one or more executors that issue the concrete tool calls. We adopt it as prior-art infrastructure rather than as a contribution: it supplies the agent scaffold, the closed toolset, and the sandboxed environment, and it validates a submitted flag against the challenge oracle. Holding this apparatus fixed is what allows the model and the prompting strategy to be isolated as the variables of interest.

The harness/framework is an \emph{instantiation}, not a dependency, of the method. \reveree consumes a normalized trajectory which is a sequence of \texttt{(thought, action, observation)} steps together with the run's validated outcome. \reveree's schema, two-channel detection, progress metrics, and behavioral profiling operate entirely on that representation. The only
harness-specific component is the adapter that produces it
(Section~\ref{sec:reveree-inputs}); a different harness, a different toolset, or
a single-agent rather than a multi-agent scaffold changes the adapter, but nothing downstream. The diagnostic methodology therefore transfers unchanged, a property that matters precisely because harnesses are evolving quickly.

\subsection{Factors of Variation}
\label{sec:pipeline-factors}

Within this fixed apparatus, we vary four factors (independently) and hold the rest constant; we state only the high-level design here and defer the concrete settings to the experimental
setup (Section~\ref{sec:eval-setup}).

\noindent\textbf{Varied.} 
The \emph{model}~(\ding{183}) is the primary axis: we sweep a set
of frontier models spanning the leading families and a capability gradient within
each. The \emph{prompting strategy}~(\ding{184}) varies the agent's reasoning
loop while leaving the harness and tools untouched, isolating the effect of
reasoning structure. Selected \emph{challenge features} (e.g., name and description)~(\ding{182}) and \emph{configurations and settings} (e.g., reasoning
effort and the budget caps)~(\ding{183}) are varied in targeted ablations.
The specific models, strategies, hyperparameter settings, and the composition of the challenge corpus are given in Section~\ref{sec:eval-setup}.

\noindent\textbf{Held fixed.} 
The execution environment and its reverse-engineering toolset~(\ding{185}) are identical across runs.

\section{\reveree: A Diagnostic Instrument for Reverse-Engineering Agents}
\label{sec:reveree}

In this section, we introduce \reveree,\footnote{\reveree---a
\emph{referee} for \emph{rev}(erse-engineering) challenges.} a process-level
instrument that scores how far an agent advanced through the reverse-engineering process 
rather than only whether it captured the flag. \reveree examines each run at three tiers, which answer successively
deeper questions. \emph{Tier~1} is the solve rate which measures whether the agent retrieved the flag. This is the standard outcome metric, which \reveree retains as the baseline that the remaining tiers explain.
\emph{Tier~2} measures progress. Specifically, it assesses how far a run advanced through the reverse engineering stages. \emph{Tier~3} measures \emph{behavior}: it views a run as a state machine over the agent's actions and extracts patterns from it.

\subsection{Design Principles}
\label{sec:reveree-principles}

Three principles separate \reveree from prompting an LLM to grade a trajectory: \textbf{1) An RE-specific fixed schema.} Credit is given based on eight canonical reverse-engineering stages, scored uniformly across challenges, rather than a generic competency rubric or hand-authored per-challenge milestones (Section~\ref{sec:reveree-schema}). \textbf{2) The right tool for each stage.} Deterministic verification is used for mechanical stages such as triage, detecting
  solver artifacts, and flag validation. Only
  comprehension stages are referred to an LLM judge
  (Section~\ref{sec:reveree-detect}). \textbf{3) Process-centered.} Beyond a partial-credit score, \reveree reports a stall funnel that localizes where runs fail and analyzes behavioral patterns of agents
  (Sections~\ref{sec:reveree-metrics},~\ref{sec:reveree-behavior}).

\subsection{Inputs}
\label{sec:reveree-inputs}

\reveree scores runs produced by any CTF agent
and consumes each run from three files: First, the \emph{agent run} which records agent output and interactions with tools.  For D-CIPHER~\cite{dcipher}, a multi-agent CTF framework, this includes a planner message list and one or more executor
sessions, \texttt{submitted\_flag} and
\texttt{success} fields. Second, a \emph{solution sketch} which is a short ground-truth description of how the challenge is solved. This is used to give the judge challenge-specific meanings for each stage but \emph{no flag}. Third, \emph{run metadata} which carries identity and configuration information.

\begin{figure}[t]
\centering
\includegraphics[width=0.8\linewidth]{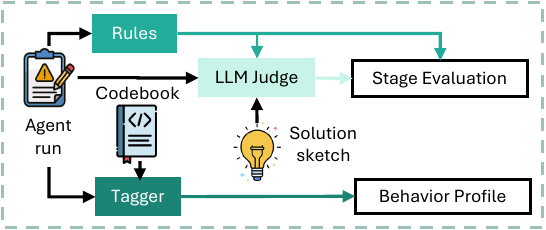}
\caption{The \reveree pipeline.}
\label{fig:reveree-arch}
\end{figure}

\subsection{The Eight-Stage RE Schema}
\label{sec:reveree-schema}

The unit of credit is the \emph{stage}, not a hand-authored per-challenge
milestone, so no manual milestone labeling is needed; the solution sketch gives challenge-specific stage context. 
\reveree fixes the RE stages based on common steps that RE analysts perform as discussed in Section~\ref{sec:bg-re}. The eight stages are derived from progression documented in observational studies of human RE analysts~\cite{votipka,bryant2012sensemaking,ceccato2019understanding}.\hadjer{[DONE]} \gang{Briefly mention/remind readers that the stages used in our paper are well aligned by prior user studies with real-world RE analysts.}
The stages are shown in
Table~\ref{tab:schema} and span the RE milestones from triage to flag validation. Each stage is tagged \textsc{action} (an externally observable operation that leaves a mechanical trace) or \textsc{insight} (an internal comprehension step), and this tag governs how the stage is detected. 
For practicality, we split the algorithm stage into
\emph{recognize} (S4a) and \emph{name} (S4b), so that spotting a transformation (e.g., a loop XORing against a constant) results in comprehension credit even when the agent never attaches the textbook name. This is a more granular measurement than a single ``identified the
algorithm'' bit. 

\begin{table}[t]
\centering
\small
\setlength{\tabcolsep}{4pt}
\resizebox{\columnwidth}{!}{
\begin{tabular}{@{}llp{0.46\columnwidth}@{}}
\toprule
\textbf{Stage} & \textbf{Type} & \textbf{What it denotes} \\
\midrule
S1 \ Triage             & \textsc{action}  & File/format, architecture, packing and protections. \\
S2 \ Surface mapping    & \textsc{action}  & Disassembly/decompilation exposing program structure. \\
S3 \ Control flow       & \textsc{insight} & The relevant branching/gating logic is understood. \\
S4a Algorithm recognize & \textsc{insight} & Core transformation spotted via a beacon (loop, constant, XOR). \\
S4b Algorithm name      & \textsc{insight} & Transformation explicitly named (``this is RC4 / base64''). \\
S5 \ Constraint extract & \textsc{insight} & The concrete conditions a valid input must satisfy are stated. \\
S6 \ Solver             & \textsc{action}  & A solver/inversion artifact is built (keygen, patch, SMT, \texttt{angr}). \\
S7 \ Flag               & \textsc{action}  & The correct flag is produced and validated. \\
\bottomrule
\end{tabular}
}
\caption{The eight-stage RE schema. \textsc{action} stages leave a mechanical
trace and are verified deterministically; \textsc{insight} stages require
demonstrated comprehension and are the only ones the LLM judge rules on. 
}
\label{tab:schema}
\end{table}

\subsection{Two-Channel Stage Detection}
\label{sec:reveree-detect}

Each stage's verdict comes from one of two channels. The principle is to use
deterministic evidence wherever the evidence \emph{is} the accomplishment, and
to invoke the judge only where the accomplishment is a matter of comprehension.

\hadjer{[DONE]}\gang{How are these rules executed? By an LLM or an author-written script? While it is deterministic, it does not guarantee correctness --- we should mention this and add a pointer to our evaluation.}

\noindent\textbf{Channel~A.} This channel is scored using manually written deterministic rules/scripts that are applied to the executed tool calls and their outputs. For each stage, the tool emits one of \texttt{confirm}, \texttt{deny},
or \texttt{abstain} outcomes. It
\texttt{confirms} where the artifact is the accomplishment, such as S7 being confirmed by
extracting the flag, S6 by a solver artifact, and S1 by a triage tool having run. Elsewhere when, for example, S2 cannot have occurred if no disassembler or decompiler was ever called, so their absence yields a deterministic negative and Channel~A outputs a \texttt{deny}. In cases where there is an uninformative invocation of a tool it (\texttt{abstains}) and defers to Channel~B. 
Channel~A also abstains on the pure-comprehension stages S3, S4a, and S4b, S5 which leave no obvious mechanical trace. The matching invoked by Channel~A uses keyword buckets for triage, disassembly/debugging, and solver tools where each is matched against the actual shell commands on whole-word boundaries, with underscores and dots treated as separators. This ensures a tool name embedded in a filename never false-fires, and merely \emph{naming} a tool or technique in plan prose (``solve,'' ``invert,'' ``patch'') cannot confirm a stage.

\noindent\textbf{Channel~B.} Channel B is the LLM judge, consulted \emph{only} for the comprehension
stages and \emph{only} where Channel~A abstains, so a run that fails early or solves mechanically may incur zero judge calls. To avoid the pitfalls of using LLM-as-a-judge~\cite{wang2024fairevaluators, zheng2023judging}, four constraints separate it from a free-form ``grade this transcript'' prompt.
It is \emph{stage-scoped}: the judge is asked only about the specific stages in
question, each defined by the solution sketch, never about generic competencies. It is \emph{outcome-blinded}: the submitted flag, the true flag, and
any correct/incorrect marker are redacted before the judge sees the trajectory,
so it cannot anchor on ``it solved, therefore it understood.''
It is \emph{evidence-bound}: every positive finding must cite a concrete
trajectory step and a short verbatim quote; if a citation is absent, the stage is not
credited. And it is
\emph{execution-grounded}: alongside the planner's reasoning, the judge receives the executor's real tool calls and outputs. This grounding lets the citation requirement be met against what the agent actually observed.

\hadjer{-----------------}

\hadjer{\textbf{[DONE]}}
\gang{Overall, this section needs a lot of examples to explain the metrics. }
\gang{I think the readers may have a hard time understanding these metrics without an example. You may consider adding an example of a agent's multiple runs: S1-S2-S3-S4 and S1-S3-S4, and then explain what each metric's value will be for the example}. 
\gang{If a metric is not used in later analysis, then we don't need to include them here.}
\subsection{Tier-2: Progress Metrics}
\label{sec:reveree-metrics}
\noindent\textbf{A worked example.} Consider a challenge whose applicable stages are S1--S7, scored for three runs:

\begin{center}
\small
\begin{tabular}{llccc}
\toprule
Run & Stages reached & Stage & Depth & Outcome \\
\midrule
A & S1, S2, S3, S4a & $4/8$ & $4/8$ & unsolved \\
B & S1, S3, S4a & $3/8$ & $1/8$ & unsolved \\
C & S1, S2, S6, S7 & $4/8$ & $2/8$ & solved \\
\bottomrule
\end{tabular}
\end{center}

\noindent The \emph{stage coverage} counts every stage reached, crediting progress regardless of gaps: Run~A reaches four of eight stages ($4/8$), Run~B three ($3/8$). \emph{Progress depth} counts only the longest unbroken prefix $S1\!\to\!S2\!\to\!\cdots$ and stops at the first miss: Run~A is intact through S4a ($4/8$), but Run~B skips S2, so its depth collapses to S1 alone ($1/8$) despite reaching S4a. When coverage exceeds depth, e.g., Run~B ($3/8$ vs.\ $1/8$), the run reached later stages without the intermediate ones.

\noindent\textbf{The stall funnel.}
\hadjer{\textbf{[DONE]}}\gang{this sentence is hard to parse.}
\gang{I don't understand what ``margional'' reach means, need an example.}
The \emph{marginal reach} of a stage is the fraction of runs that reached \emph{that} stage not the fraction that reached at least that far. Each cell is that fraction within a population: for stage~S and population~P, it is the number of runs in~P that reached~S divided by $|P|$. Taking the S2 column as a worked cell: two of the three runs (A and C) reached S2, so the ``All'' entry is $2/3$; of the runs, only C is solved and it reached S2, giving $1/1$ for ``S''; among the two unsolved runs only A reached S2, giving $1/2$ for ``U''.
Applying this to every stage yields:
\begin{center}
\small
\begin{tabular}{lcccccccc}
\toprule
 & S1 & S2 & S3 & S4a & S4b & S5 & S6 & S7 \\
\midrule
All     & $3/3$ & $2/3$ & $2/3$ & $2/3$ & $0/3$ & $0/3$ & $1/3$ & $1/3$ \\
S   & $1/1$ & $1/1$ & $0/1$ & $0/1$ & $0/1$ & $0/1$ & $1/1$ & $1/1$ \\
U & $2/2$ & $1/2$ & $2/2$ & $2/2$ & $0/2$ & $0/2$ & $0/2$ & $0/2$ \\
\bottomrule
\end{tabular}
\end{center}
A cumulative curve would read $1$ at every stage for the solved run C (because it reached S7), hiding that C skipped S3--S5. The marginal view exposes this.
The ``solved-without-understanding'' or ``shortcut solve'' signature of a flag produced without the underlying analysis, while the unsolved row shows the opposite (comprehension reached, no solver or flag). 
Table~\ref{tab:funnel_bysolve} shows this contrast at corpus scale on picoCTF.
\hadjer{\textbf{[DONE]}}\gang{Need a walk-through example to understand how this metric captures the ``solved-without-understanding''. }

\subsection{Tier-3: Behavioral Profiling}
\label{sec:reveree-behavior}

The milestone tiers characterize \emph{what} an agent accomplished; they do not clarify \emph{how} it spent its effort. This procedural dimension is central
to reverse engineering, where progress typically depends on \emph{pivoting}
between static inspection, dynamic observation, and hypothesis-driven scripting
rather than exhausting any one in
isolation~\cite{votipka2020observational,ceccato2019understanding}. The Tier~3
represents each run as a sequence of behavioral states and summarizes it with a
small set of pre-determined and interpretable features. 

\noindent\textbf{State set and tagging.} Each action is classified at two
granularities (Table~\ref{tab:states}): a six-state coarse states over which we
estimate a Markov model: \textsc{Triage}, \textsc{Static} (disassembly,
decompilation), \textsc{Dynamic} (debugging, tracing, execution), \textsc{Script}
(building or running a solver), \textsc{Validate} (flag submission), and
\textsc{Other} and a more granular layer of 20 codes that are categorized under these states. The 20 fine-grained behavioral codes were developed by iterative coding of a pilot set of trajectories, then grouped under the six coarse states.
\hadjer{[DONE]}\gang{How are the codes created? Based on what?}
A precedence-ordered, command-pattern tagger maps each action to a (fine, coarse)
pair, first match winning, so a compound shell line is assigned a single dominant
intent; the ordering encodes specificity, so ``\texttt{gdb ./bin}'' is
\textsc{Dynamic}, ``\texttt{file ./bin}'' is \textsc{Triage}, and a bare
``\texttt{./bin}'' is \textsc{Dynamic} only when no analysis tool has already claimed
the command. The harness's tools (decompiler, disassembler,
file-authoring) are mapped directly to \textsc{Static}, \textsc{Static}, and
\textsc{Script}, so analysis through dedicated tools is not misattributed to
\textsc{Other}; because the action space is a closed toolset. This results in unclassified
residual and an unseen shell-command strings accounting to under $1\%$ of actions on our
corpus. 

\noindent\textbf{Per-run representation.} 
We summarize each run as a first-order Markov chain over the six coarse interaction states in $\mathcal{S}$. Given a run's state sequence $s_1,\dots,s_T$, we count transitions only \emph{within} a session.
$n_{ij}$ denotes the resulting within-session count of $i\!\to\!j$ transitions. From this matrix and the tagged action stream we read a small feature vector, each component capturing one behavior of interest. Refer to Appendix~\ref{sec:markov-matrix} for a detailed explanation of each metric. 


\hadjer{[time-permitting]}\gang{a lot of metrics are defined, but these are relatively easier to understand. Maybe it can benefit readers to list each metric and its definitions. If the table gets too big, we can put it to the appendix.}


\begin{table}[t]
\centering
\small
\begin{tabular}{@{}ll@{}}
\toprule
\textbf{State} & \textbf{Meaning (representative tools)} \\
\midrule
\textsc{Triage}   & File/format reconnaissance (\texttt{file}, \texttt{strings}). \\
\textsc{Static}   & Disassembly / decompilation (r2, \texttt{objdump}). \\
\textsc{Dynamic}  & Debugging, tracing, executing file (\texttt{gdb}, \texttt{ltrace}). \\
\textsc{Script}   & Building/running script (Python, z3/\texttt{angr}, patching). \\
\textsc{Validate} & flag submission. \\
\textsc{Other}    & Navigation, file reads, environment setup. \\
\bottomrule
\end{tabular}
\caption{The six coarse behavioral states. The tagger
assigns a pair of coarse fine-grained codes.}
\label{tab:states}
\end{table}

\gang{------------------------}

\hadjer{\textbf{[DONE]}}

\gang{I think readers will cast doubt on the reliability of LLM judges given the rich body of literature on this. Unfortunately, there is no other way to reliably evaluate ``comprehension''. The above discussion of ``grounding'' helps a lot. Ultimately, it is still up to the evaluation.  }
\gang{The problem I see is that the evaluation (Section 5) directly dives into evaluating frontier models. However, there is no evaluation on Reveree itself. Channel A is deterministic, but the methodology may not guarantee correctness of judging --- we should evaluate it. Channel B is an LLM judge which should be evaluated given the prior works and concerns. We can put the evaluation in Appendix if we run out of space. I recall that we manually labeled 15 challenges --- we can report \% of correctly judged stages and LLM judge performance on these 15 challenges to make reviewers believe the results are reliable. 
}

\subsection{Judge Validation}
\label{sec:judge-validation}
\reveree's scoring is deterministic when the presence of the evidence \emph{is} the
accomplishment. This applies to S1, S6, and S7 (decided by Channel~A rules), and S2 (which is decided by Channel~A when the tool output contains structural
markers, and falls back to the judge otherwise). The LLM judge is
consulted for the comprehension/insight stages S3, S4a, S4b, and S5: the process is outcome-blinded, and must cite a trajectory span for every
verdict (Section~\ref{sec:reveree-detect}). To ensure the LLM Judge outcome is reliable, we validate the pipeline by measuring whether the judge's comprehension verdicts match those of a human expert. Additionally, we audit the deterministic rules for false positives and negatives.

\noindent {\bf Human agreement study.}
We sampled 15 runs from the judged corpus, chosen at random subject
to covering both benchmarks (8 NYU-CTF, 7 picoCTF), both outcomes
(9 solved, 6 unsolved), and the study's two model families (5
Claude-family, 6 GPT-family, and 4 Codex runs, spanning six
models); the sample covers 12 unique challenges, three of which
appear twice under different agents
(Appendix~\ref{app:validation-sample}, Table~\ref{tab:validation-sample}).
An author with eight years of
reverse-engineering experience served as the annotator and independently labeled all eight stages for each run yielding 120 (run, stage) decisions. The coding scheme was the eight-stage schema
itself: the annotator applied the stage definitions of
Section~\ref{sec:reveree-schema} directly, and read the same inputs the judge receives: the agent run and the per-challenge solution brief. For S1, S2, S6, and S7, the human labels audit Channel~A;
for S3--S5 the labels are compared against the judge's Channel~B verdicts.

\noindent {\bf Results.}
Table~\ref{tab:judge-human} reports per-stage agreement. Overall
agreement is 94.2\% (113/120; 95\% CI [88.4, 97.1]). On the
deterministic stages the audit found no rule errors (60/60; pooled CI
[94.0, 100]). On the judged comprehension stages, judge--human
agreement is 88.3\% (53/60; CI [77.8, 94.2]). For the insight stages it is: 93.3\% for control-flow comprehension (S3) and algorithm recognition (S4a), 86.7\% for algorithm naming (S4b), and 80.0\% for constraint extraction (S5). All seven disagreements fall in the judged stages.

\begin{table}[t]
  \centering
  \small
  \begin{tabular}{llrll}
    \toprule
    Stage & Agree & Rate & 95\% CI \\
    \midrule
    S1              & 15/15 & 100.0\% & [79.6, 100] \\
    S2        & 15/15 & 100.0\% & [79.6, 100] \\
    S3             & 14/15 & 93.3\%  & [70.2, 98.8] \\
    S4a         & 14/15 & 93.3\%  & [70.2, 98.8] \\
    S4b            & 13/15 & 86.7\%  & [62.1, 96.3] \\
    S5             & 12/15 & 80.0\%  & [54.8, 93.0] \\
    S6           & 15/15 & 100.0\% & [79.6, 100] \\
    S7              & 15/15 & 100.0\% & [79.6, 100] \\
    \midrule
    Judged     & 53/60 & 88.3\%  & [77.8, 94.2] \\
    Deterministic     & 60/60 & 100.0\% & [94.0, 100] \\
    Overall                     & 113/120 & 94.2\% & [88.4, 97.1] \\
    \bottomrule
  \end{tabular}
  \caption{Human agreement with \reveree's per-stage verdicts on 15 runs (120 decisions). S1, S2, S6, and S7 audit the Channel~A rules.
  S3--S5 are compared against the Channel-B LLM judge. Intervals are Wilson score intervals.}
  \label{tab:judge-human}
\end{table}

\noindent {\bf Disagreement analysis.}
All seven disagreements fall in the judged stages, and they are
concentrated rather than scattered. 
For example, a single solved run accounts for four of the seven. In this run, the annotator credited S3, S4a, S4b, and S5 while the judge credited none of them. Note that six of the seven disagreements happened when {\em the LLM judge was too conservative}: human annotator marked the stage reached while judge did not. The single LLM judge false positive is an S4b verdict on a solved run where the judge accepted an algorithm naming while the annotator did not. 

Overall, the result confirms that \reveree provides reliable evaluations. 
\gang{I added a conclusion remark.}

\makeatletter
\@ifpackageloaded{tcolorbox}{%
  \definecolor{insightteal}{rgb}{0.059,0.463,0.431}%
  \newtcolorbox{insightcallout}{colback=insightteal!6, colframe=insightteal,
    boxrule=0.6pt, arc=2.5pt, left=8pt, right=8pt, top=5pt, bottom=5pt,
    before upper={\textbf{\textsc{\textcolor{insightteal}{Insight.}}}\hspace{0.5em}}}%
  \newcommand{\insight}[1]{\begin{insightcallout}\small #1\end{insightcallout}}%
}{%
  \newcommand{\insight}[1]{\par\medskip\noindent
    \fbox{\begin{minipage}{\dimexpr\columnwidth-2\fboxsep-2\fboxrule\relax}%
      \small\textbf{\textsc{Insight.}}\hspace{0.5em}#1\end{minipage}}\par\medskip}%
}
\makeatother
\graphicspath{{figs_merged/}}

\section{Evaluation}
\label{sec:eval}
In this section, we evaluate state-of-the-art LLM agents on reverse-engineering (RE) capture-the-flag challenges. We organize the evaluation around three research questions:

\begin{enumerate}[leftmargin=*,label=\bfseries
  RQ\arabic*:,
  ]
\item How does the performance of an LLM agent vary across frontier models, prompting strategies, and configurations on RE tasks? 
\item What causes LLM agents to fail on RE tasks, and which behavioral signatures distinguish failure modes? 
\item Do LLM agents solve RE challenges through memorization and surface-features, or genuine
  analysis?
\end{enumerate}

\subsection{Experimental Setup}
\label{sec:eval-setup}

\textbf{Models.} We evaluate nine frontier models from the two leading families, spanning different tiers/sizes and versions within each family
(Table~\ref{tab:modelset}) while holding the agent framework
(D-CIPHER~\cite{dcipher}) fixed. Codex is a code-specialised
variant included as a point of contrast.
 
\begin{table}[h]
\centering
\small
\begin{tabular}{lccc}
\toprule
Model & Family & Size & Version \\
\midrule
haiku    & Claude & Small  & 4.5 \\
sonnet   & Claude & Medium & 4.5 \\
opus 4.5 & Claude & Large  & 4.5 \\
opus 4.7     & Claude & Large  & 4.7 \\
\midrule
nano      & gpt5  & Small  & 5.0 \\
mini      & gpt5  & Medium & 5.0\\
gpt5           & gpt5  & Large  & 5.0 \\
gpt5.4         & gpt5  & Large  & 5.4 \\
codex           & gpt5  & Large & 5.3\\
\bottomrule
\end{tabular}
\caption{The nine evaluated models and the axes varied within each family. 
\emph{Size} is the capability tier and \emph{version} is the model release.}
\label{tab:modelset}
\end{table}

\noindent\textbf{Prompting strategies.} We compare four widely used prompting
strategies, holding the harness and tools fixed so that only the prompt differs.
\emph{Vanilla}, the default D-CIPHER prompt left unchanged: the agent is given the task and the available tools and acts directly, with no imposed reasoning format and no cross-step memory (baseline). \emph{Ledger} carries a persistent scratchpad of state across steps: the planner is instructed to maintain a structured ledger of established facts and open hypotheses in its reasoning output~\cite{scratchpad}, but no harness-level enforcement is imposed. \emph{ReAct}~\cite{react} interleaves explicit reasoning with
actions: the agent alternates between a natural-language thought about the current state and a tool call, so each action is conditioned on the preceding reasoning and each observation feeds the next thought. \emph{Reflexion}~\cite{reflexion} adds a verbal self-critique after a failed attempt: the agent reflects on why the attempt failed and carries that reflection forward as guidance for the next try. The four \emph{multi-strategy} models (opus~\cite{claude}, codex~\cite{codex}, gpt5~\cite{gpt}, gpt5.4~\cite{gpt}) are run under all four strategies; the remaining five are run under vanilla only.

\begin{figure*}[t]
\centering
\begin{subfigure}[t]{0.38\linewidth}
  \centering
  \includegraphics[width=\linewidth]{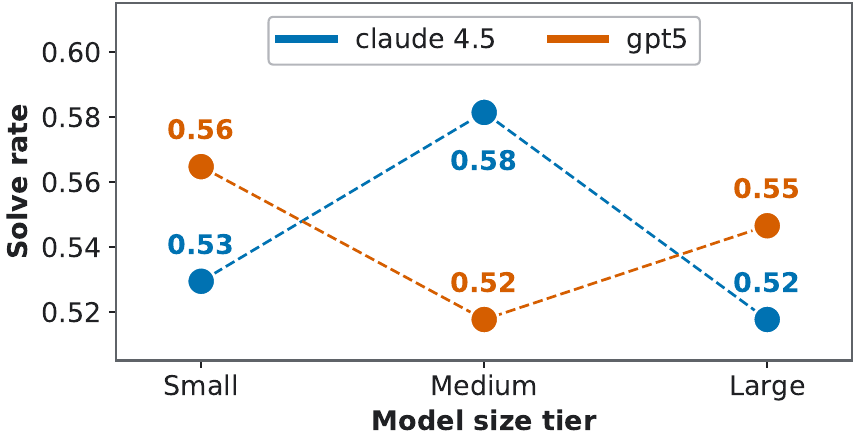}
  \caption{Solve rate vs.\ model size tier.}
  \label{fig:sizever_a}
\end{subfigure}
\begin{subfigure}[t]{0.38\linewidth}
  \centering
  \includegraphics[width=\linewidth]{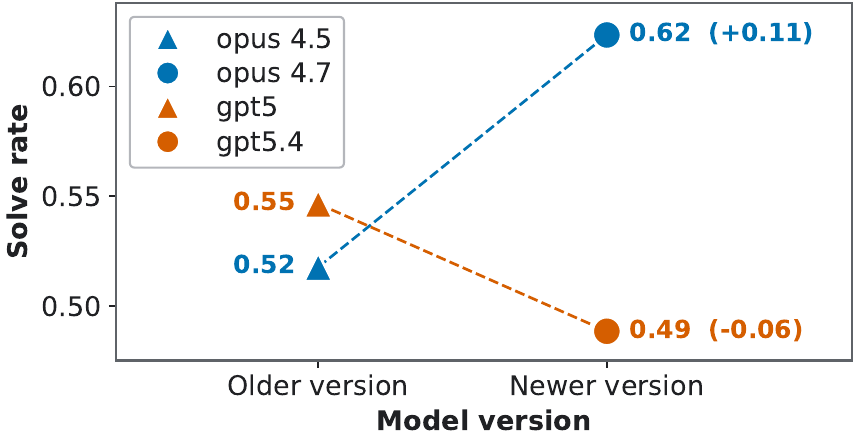}
  \caption{Solve rate across version updates.}
  \label{fig:sizever_b}
\end{subfigure}
\caption{Within-family model comparison
\textbf{(a)}~Mean solve rate for each family (claude 4.5 in blue, gpt5 in orange) across three size tiers (small/medium/large).
\textbf{(b)}~Mean solve rate of each family's older vs.\ newer version (marker shape denotes version), with the per-model change annotated. Solve rate is the fraction of runs reaching stage~S7, pooled over the full dataset.}
\label{fig:sizever}
\end{figure*}

\noindent\textbf{Challenges.} We evaluate on 88 RE challenges from two sources. Intercode-CTF~\cite{yang2023intercode} which collects challenges from PicoCTF and NYU-CTF~\cite{nyuctf} and they contribute 27 and 61 challenges, repectively.

\noindent\textbf{Scoring with \reveree.} Each trajectory is scored by \reveree against an eight-stage RE schema (Section~\ref{sec:reveree-schema}).
We report three milestone metrics: \emph{solve rate}, the fraction of runs reaching S7; \emph{stage coverage}, 
\hadjer{done, changed name to "stage coverage" to be more intuitive}\gang{I forgot what the macro is even though I just read the metrics section in Section 4. I think we call it existence score? We should be consistent with the naming. } the fraction of stages a run
reaches
which grants partial credit for how far the RE process got;
\hadjer{done:removed}\gang{how is it weighted? Again, need to clarify earier in Section 4}
and \emph{progress depth}, the longest unbroken chain of consecutive stages.
\gang{We can remove this if not reported or discussed.}
Additionally, we extract each run's
\emph{exit reason} directly from the termination log.
We also generate a vector of behavioral features over six coarse states 
from \reveree's Tier-3 (Section~\ref{sec:reveree-behavior}).

\gang{Overall comments for the evaluation results: move figures and tables closer to the text describing them (applying to all figures and tables.)}

\subsection{RQ1: Performance Across Configurations}
\label{sec:eval-rq1}

\subsubsection{Models}
\label{sec:eval-rq1-models}

Table~\ref{tab:models} reports per-model performance under vanilla prompting.
The ranking by solve rate and the ranking by partial progress (stage coverage) disagree. Opus 4.7 has the highest solve rate (0.624). Codex, in contrast, has the
\emph{lowest} solve rate (0.447) but the \emph{highest} stage coverage (0.621). Additionally, at
\$0.86 per solve it is roughly $2.5\times$ cheaper than any other model. Leveraging the milestone schema used by \reveree (Section~\ref{sec:reveree-schema}), this means codex advances furthest through the RE process but converts that progress into a captured flag least often. A manual review of a subsample of codex's trajectories corroborates the pattern: on the cases we examined, codex carries out the mechanical steps of RE well. It generates working scripts and uses the analysis tools effectively, but stalls before turning that work into a flag. 
The profile is consistent with a code-specialized model: strong at the
individual operations RE requires (e.g., calling tools, writing, and running scripts), but is not able to reliably integrate their results into a validated flag. We offer this as an interpretation of the trajectories we examined rather than an established property of code-specialized models. Models also differ in how they work, e.g., tool reliance and reanalysis; Appendix~\ref{app:beh-models} reports these per-model fingerprints.
\hadjer{added comment}\gang{Do we have a reference for this type of behavior for code models, or is this our speculation for Codex?}

\begin{table}[t]
\centering
\small
\begin{tabular}{lccc}
\toprule
Model & Solve & Stage & \$/solve \\
\midrule
opus 4.7     & \textbf{0.624} & 0.563 & 2.11 \\
sonnet   & 0.581 & 0.613 & 2.63 \\
nano      & 0.565 & 0.600 & 2.39 \\
gpt5           & 0.547 & 0.587 & 2.78 \\
haiku    & 0.529 & 0.591 & 2.52 \\
opus 4.5 & 0.518 & 0.535 & 2.40 \\
mini      & 0.518 & 0.568 & 2.59 \\
gpt5.4         & 0.488 & 0.562 & 2.87 \\
codex           & 0.447 & \textbf{0.621} & \textbf{0.86} \\
\bottomrule
\end{tabular}
\caption{Per-model performance under vanilla prompting, sorted by solve rate.
Cost is USD (\$) per solved challenge. Bold marks the best value in each column.}
\label{tab:models}
\end{table}


\begin{figure*}[t]
\centering
\begin{subfigure}[t]{0.32\linewidth}
  \centering
  \includegraphics[width=\linewidth]{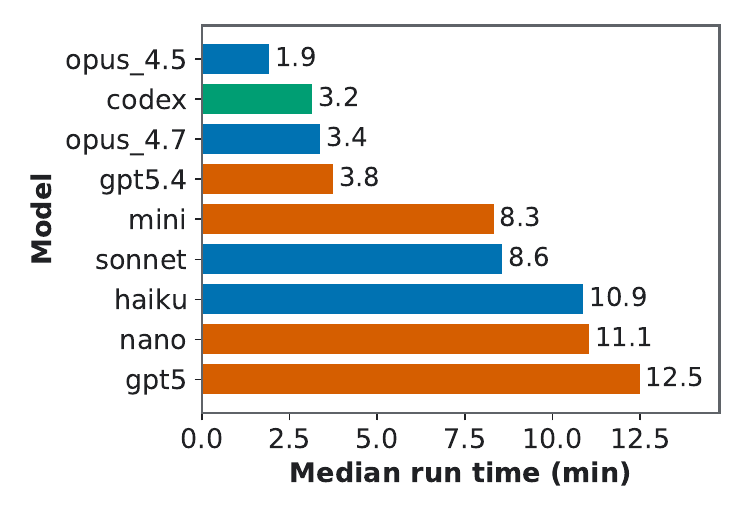}
  \caption{Per-model run time.}

  \label{fig:time_a}
\end{subfigure}
\hfill
\begin{subfigure}[t]{0.32\linewidth}
  \centering
  \includegraphics[width=\linewidth]{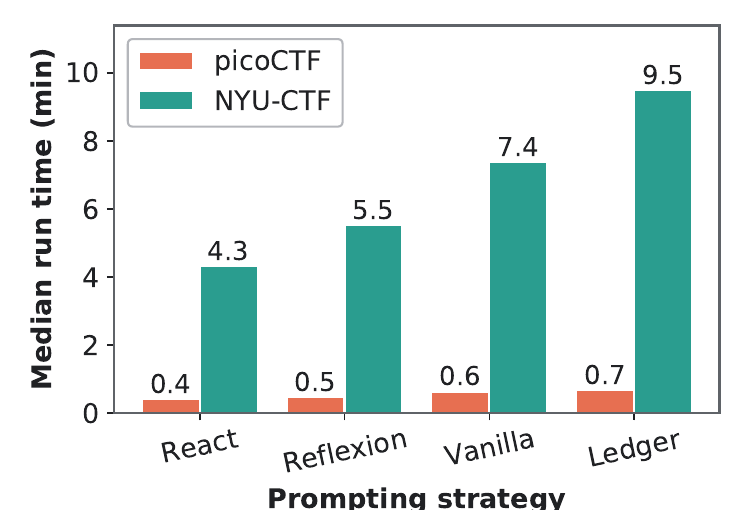}
  \caption{Per-strategy run time.}
  \label{fig:time_b}
\end{subfigure}
\hfill
\begin{subfigure}[t]{0.32\linewidth}
  \centering
  \includegraphics[width=\linewidth]{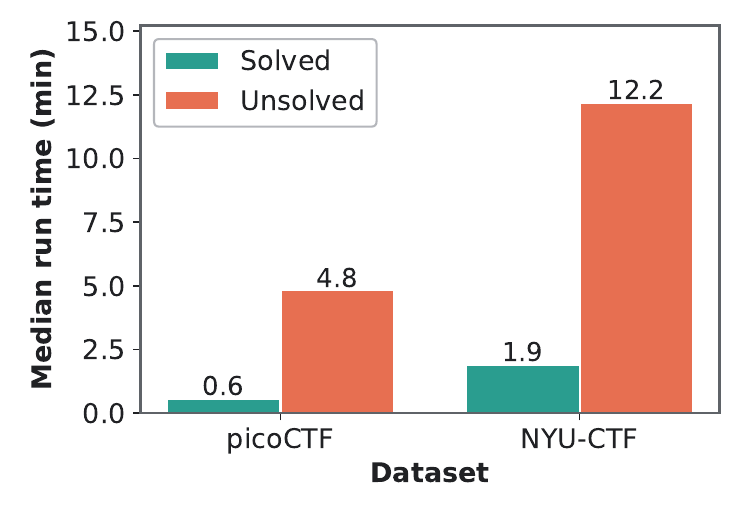}
  \caption{Solved vs.\ unsolved by dataset.}
  \label{fig:time_c}
\end{subfigure}
\caption{Wall-clock run time.
\textbf{(a)}~Median run time (minutes) per model on the \texttt{vanilla} strategy.
\textbf{(b)}~Median run time per prompting strategy, over the four multi-strategy models.
\textbf{(c)}~Median run time for solved vs.\ unsolved runs, grouped by dataset. 
\hadjer{\textbf{[DONE]}}
\gang{The numbers placed on top of the bars are still too small to read.} 
\gang{mini should be gpt5mini. } 
\hadjer{[across all the paper I use mini, nano, sonnet, without family name I also do not capitalize, do you think it would be better to change that?]}
}
\label{fig:time}
\end{figure*}

\begin{figure*}[t]
\centering
\begin{subfigure}[t]{0.26\linewidth}
  \centering
  \includegraphics[width=\linewidth]{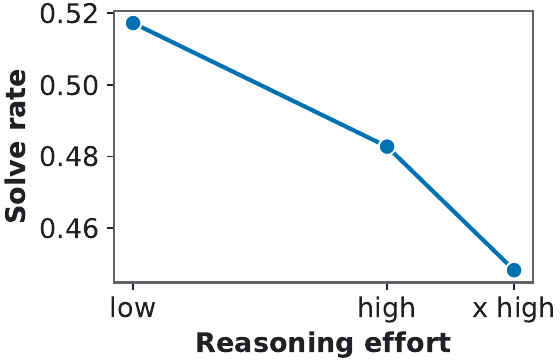}
  \caption{opus 4.7}
  \label{fig:reason_a}
\end{subfigure}
\begin{subfigure}[t]{0.26\linewidth}
  \centering
  \includegraphics[width=\linewidth]{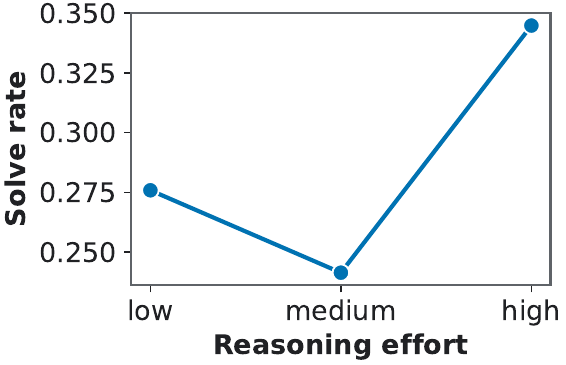}
  \caption{codex}
  \label{fig:reason_b}
\end{subfigure}
\begin{subfigure}[t]{0.26\linewidth}
  \centering
  \includegraphics[width=\linewidth]{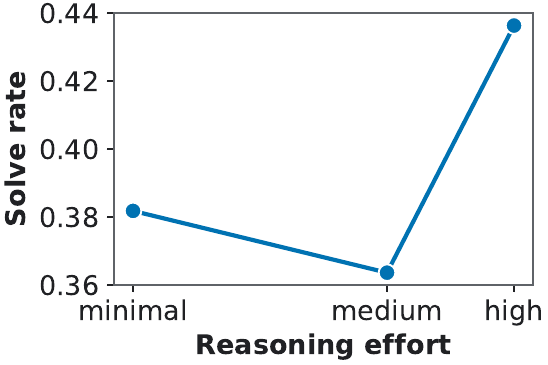}
  \caption{gpt5}
  \label{fig:reason_c}
\end{subfigure}
\caption{
Each panel plots solve rate on NYU-CTF  challenges across that model's reasoning-effort levels.
}
\label{fig:reason}
\end{figure*}

\noindent\textbf{Model size.} 
\hadjer{added a clearer footnote}\gang{I don't understand the logic here. Why is Opus 4.7 not good for the size ladder? This decision is not very big deal but the logic needs to be more clear.}\hadjer{bcs version 4.7 does not have sonnet - the medium tier}
Neither family scales monotonically. Using a small\,/\,medium\,/\,large ladder of haiku\,/\,sonnet\,/\,opus~4.5~\footnote{For this experiment we use  v4.5 because a size comparison must hold the version fixed, and the v4.7 has no medium (sonnet) or small (haiku) counterparts. For the other experiments, we use the stronger opus~4.7 as our main model for the Claude family.} (Fig.~\ref{fig:sizever_a}), Claude peaks
at the \emph{medium} tier: sonnet ($0.581$) is the strongest,
and the large opus 4.5 ($0.518$) drops below it.
This means that the largest model is the \emph{weakest}. For gpt5: nano ($0.565$) matches/surpasses the full gpt5
($0.547$), mini is the weakest of the three ($0.518$).
These results show that within a version, size does not buy RE skill for either family.

\noindent\textbf{Model versions.} Fig.~\ref{fig:sizever_b} shows that a newer model is not dependably a stronger one. The opus model has a large gain comparing version 4.5 to 4.7
($+0.106$ solve overall, $+0.138$ on NYU-CTF).
In the other direction, gpt5 outperforms gpt5.4 ($-0.058$ overall, $-0.051$ in NYU-CTF). We use \reveree to localize the source of the regression
and highlight two changes: First, gpt5.4 exhausts the budget cap on a larger
share of its failures (73\% vs. 69\%) and reaches that cap in fewer executor
turns (129 vs. 177) at comparable total cost. This translates to a higher per-step cost leading to less exploration. Second, its failures shift away from the round limit toward giving up (24\% vs. 2\%).


\subsubsection{Time Analysis}

\textbf{Model speed.} A run's
wall-clock time is set by how long the agent persists before it
terminates (by solving, by giving up, or by exhausting the budget or round
cap).
The quickest-finishing configurations are opus 4.5 ($1.9$\,min
median), codex ($3.2$ min), opus 4.7 ($3.4$ min), and
gpt5.4 ($3.8$ min): a set that spans the entire solve-rate range, from the strongest solver (opus 4.7) to the weakest (codex), and that
is dominated by large opus models rather than small ones. What unites
them is decisiveness
not accuracy or size. The
slowest runs come from gpt5 ($12.5$ min), nano ($11.1$ min), and
 haiku ($10.9$ min), a group of mixed size that includes the
full-scale gpt5; these 
because they \emph{grind}, making disorganized efforts to reach the solution (Fig.~\ref{fig:time_a}). Counter to the intuition that smaller models are faster, the quickest agents here are the largest.
 
\noindent\textbf{Solves vs. time relationship.} When an agent solves a challenge, it exits faster than when it is unable to solve a challenge. Median time-to-solve is under a minute for the strongest
configurations. Across all models, $70\%$ of solves land within $2$ minutes and
$85\%$ within $5$ minutes. The wall-clock is therefore spent overwhelmingly on failures,
which run $6$--$8\times$ longer than solves (Fig.~\ref{fig:time_c}). On
NYU-CTF, a solved run takes a median $1.9$\,min against $12.2$\,min for an
unsolved one. PicoCTF has a similar pattern solved runs take a median of $0.6$\,min against $4.8$\,min for an
unsolved one. 


\noindent\textbf{Prompting strategy speed.} As shown in Fig.~\ref{fig:time_b}, the
reasoning-led \texttt{ReAct} ($2.2$\,min) and \texttt{Reflexion} ($2.6$\,min) are fastest and
the memory-heavy \texttt{Ledger} ($6.1$\,min) slowest, with \texttt{Vanilla} ($4.1$\,min)
between them; but the difference is small on the easier picoCTF (all $0.4$--$0.7$\,min) and is more noticeable on NYU-CTF ($4.3$--$9.5$\,min). \hadjer{paraphrased}\gang{This sentence is hard to understand, what do you mean by ``this spread is minute?''}
The dataset gap is an order of magnitude \emph{within every strategy} ($10$--$14\times$).



\subsubsection{Prompting Strategies}
\label{sec:eval-rq1-strat}
Whether a prompting strategy helps depends on how hard the target is. On the
saturated picoCTF the choice is immaterial: all four strategies solve between
0.852 and 0.880, a 2.8-point spread 
(Table~\ref{tab:modelstrat}). On the harder
NYU-CTF the two reasoning-based strategies perform better. In the best performing model, opus, ReAct reaches
0.5 and Reflexion 0.552. This is a gain of $+3.5\%$ and $+14\%$ over vanilla. Other models show a performance gain
between $+19\%$ and $+35\%$ over vanilla. Between the two reasoning-based strategies it becomes model-dependent 
(Table~\ref{tab:modelstrat}): ReAct shows better results for the gpt family (codex 0.362, gpt5 0.466,
gpt5.4 0.458), while ledger and Reflexion (both 0.552) are the highest performers for opus. These experiments show that for more difficult challenges reasoning-based prompting strategies perform better but no single strategy dominates. The strategies that help do so by changing what the agent does, validating more often and re-analyzing the binary less (Appendix~\ref{app:beh-strategies}); Section~\ref{sec:eval-rq2-behav} makes this precise at the level of success and failure.

\begin{table}[t]
\centering
\small
\setlength{\tabcolsep}{3.5pt}
\resizebox{\linewidth}{!}{
\begin{tabular}{lcccc}
\toprule
Model & Vanilla & Ledger & ReAct & Reflexion \\
\midrule
\multicolumn{5}{l}{\textit{picoCTF}}\\
opus 4.7 & \textbf{0.926} / 0.593 & \textbf{0.926} / 0.644 & \textbf{0.926} / 0.542 & \textbf{0.926} / 0.532 \\
codex       & \textbf{0.815} / 0.708 & \textbf{0.815} / 0.662 & 0.778 / 0.718 & 0.778 / 0.685 \\
gpt5       & 0.889 / 0.694 & \textbf{0.926} / 0.579 & 0.889 / 0.611 & 0.889 / 0.620 \\
gpt5.4     & 0.815 / 0.648 & \textbf{0.852} / 0.620 & 0.815 / 0.708 & 0.815 / 0.722 \\
\midrule
\multicolumn{5}{l}{\textit{NYU-CTF}}\\
opus 4.7 & 0.483 / 0.550 & \textbf{0.552} / 0.623 & 0.500 / 0.567 & \textbf{0.552} / 0.586 \\
codex       & 0.276 / 0.580 & 0.310 / 0.547 & \textbf{0.362} / 0.573 & 0.316 / 0.570 \\
gpt5       & 0.390 / 0.538 & 0.339 / 0.492 & \textbf{0.466} / 0.627 & 0.431 / 0.623 \\
gpt5.4     & 0.339 / 0.523 & 0.271 / 0.483 & \textbf{0.458} / 0.714 & 0.373 / 0.631 \\
\bottomrule
\end{tabular}
}
\caption{Solve rate/stage coverage for the four multi-strategy models under each prompting
strategy, \emph{per dataset}. {\bf Bold} marks each model's best solve rate.
}
\label{tab:modelstrat}
\end{table}


\noindent\textbf{Prompting strategy solve overlap.} Strategies solve
mostly the same cases: out of  the 
cases
solved by \emph{any} strategy, 72\% are solved by all four and only 10\% by
exactly one. Counting unique contributions, ReAct solves 10 cases that no other
strategy solves, against five for ledger, five for Reflexion, and two for
Vanilla. This shows that ReAct is not only the best on average (for NYU-CTF) but also responsible for the most unique solves.


\subsubsection{Reasoning Effort}
\label{sec:eval-rq1-config}

Finally we vary the agent's reasoning-effort setting on a matched subset of
challenges (Fig.~\ref{fig:reason}). \textit{The effect is non-monotonic and
model-specific.} Opus declines monotonically as effort increases: 0.517
with reasoning low, 0.483 at high (default setting),  and 0.448 at
extra-high. On this subset, the best results are achieved when reasoning effort is set to low. Codex and gpt5 are instead U-shaped: the medium setting is
\emph{worse} than the baseline and only the highest setting improves on it
(codex from 0.241 to 0.345, gpt5 from 0.364 to 0.436). 


\insight{The base model is the primary determinant of performance. Within a family, neither a larger size tier nor a newer version reliably improves RE performance, and neither higher cost per solve nor longer run time indicates more thorough analysis; long runs are failed runs that persist without progress. Additionally, increasing reasoning effort is not monotonically beneficial. Prompting benefits are dataset-dependent: negligible on easy challenges and beneficial on harder ones.
}


 

\subsection{RQ2: Failure Analysis and Signatures}
\label{sec:eval-rq2}

RQ1 establishes \emph{how often} agents succeed. We now use \reveree to ask
\emph{why} and \emph{where} they fail, \emph{what} separates a successful run
from a failed one, and whether targeted interventions help.

\subsubsection{Failure Modes}
\label{sec:eval-rq2-fail}

\begin{figure}[t]
\centering
\includegraphics[width=0.84\linewidth]{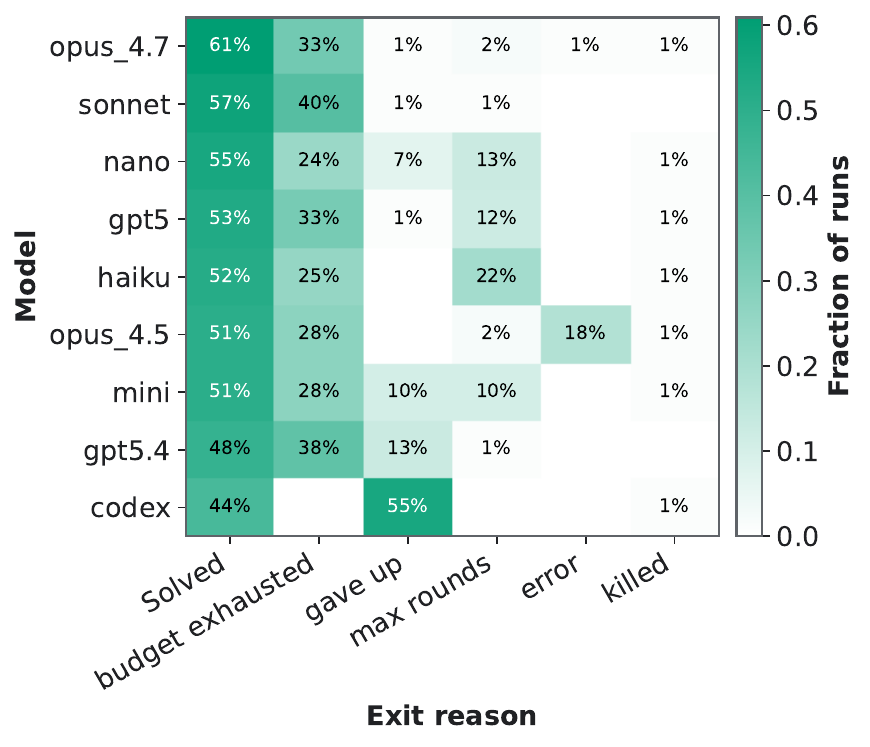}
\caption{Exit-reason composition per model, pooled across
all challenges. Each cell reports the fraction of that model's runs ending in that exit reason.}
\label{fig:exitmodel}
\end{figure}

\noindent\textbf{Why agents fail.}  Across all failed runs ($n{=}852$), the most common termination is budget exhaustion ($50\%$), followed by giving up ($25\%$), hitting the planner-round cap ($19\%$), tool
errors ($4\%$), and being killed/timeout ($2\%$). The composition differs by model (Fig.~\ref{fig:exitmodel}). 
Codex \emph{gives up} in 98\% of its failures, i.e., it stops rather than exhausting its budget. The Claude models are limited by budget instead (opus 85\%, sonnet 95\%), haiku hits the planner round cap in 45\% of its failures, and opus~4.5 is the most error-prone (37\%). The gpt5 models split between budget and the planner round cap (gpt5 69\% and 26\%) or budget and giving up (gpt5.4 73\% and 24\%).

\begin{table}[t]
\centering
\small
\begin{tabular}{lcc}
\toprule
Configuration & Rescue & $n$ (failures) \\
\midrule
codex $+$ persist                       & 0.15 & 47 \\
codex $+$ persist $+$ no giveup                     & 0.10 & 41 \\
opus 4.7 / budget & 0.14 & 28 \\
gpt5 / budget       & 0.10 & 29 \\
gpt5 / rounds          & 0.11 & 9  \\
\bottomrule
\end{tabular}
\caption{Rescue rate: fraction of same-strategy baseline failures of the targeted exit type solved after
enabling each knob.
}
\label{tab:rescue}
\end{table}

\begin{table*}[t]
\centering
\footnotesize
\begin{tabular}{ll cccccccccc}
\toprule
Dataset & Outcome & Depth & Stage & S1 & S2 & S3 & S4a & S4b & S5 & S6 & S7 \\
\midrule
\multirow{2}{*}{picoCTF} & solved   & 0.160 & 0.650 & 0.99 & 0.06 & 0.67 & 0.77 & 0.41 & 0.59 & 0.71 & 1.00 \\
                          & unsolved & 0.329 & 0.693 & 1.00 & 0.36 & 0.82 & 0.95 & 0.69 & 0.80 & 0.92 & 0.00 \\
\midrule
\multirow{2}{*}{NYU-CTF} & solved   & 0.387 & 0.710 & 1.00 & 0.64 & 0.54 & 0.76 & 0.44 & 0.49 & 0.81 & 1.00 \\
                          & unsolved & 0.362 & 0.471 & 0.99 & 0.82 & 0.54 & 0.46 & 0.21 & 0.19 & 0.54 & 0.00 \\
\bottomrule
\end{tabular}
\caption{split by outcome: RE-stage reach for solved vs.\ unsolved runs, per dataset.}
\label{tab:funnel_bysolve}
\end{table*}

\noindent\textbf{Where agents fail.} Splitting each dataset's funnel into solved versus unsolved runs (Table ~\ref{tab:funnel_bysolve}) localizes the stage where the two diverge, i.e., the point at which 
an agent fails. On the NYU-CTF dataset, the two trajectories are similar through triage, surface mapping, and control-flow recovery (S1--S3) and then split at \emph{algorithm identification} (S4a), reached by $0.76$ of solved runs but only $0.46$ 
of unsolved ones. Solved runs continue to surpass unsolved runs through constraint extraction (S5: $0.49$ vs.\ $0.19$) and solving (S6: $0.81$ vs.\ $0.54$). The bottleneck is therefore \emph{comprehension}, not early
reconnaissance or final scripting: failed agents recover the high-level
structure of the binary but cannot \emph{comprehend} the details needed to drive it to a flag. The picoCTF dataset inverts this trend: the unsolved line sits \emph{above} the solved one at nearly every analysis stage (S4a $0.95$ vs.\ $0.77$, S6 $0.92$ vs.\ $0.71$).
This might be happening because picoCTF is an easier benchmark where most challenges are solved without a full reverse (skipping stages) of the binary. This is further corroborated with low progress depth for solved challenges ($0.16$). This is also consistent with our finding that models reproduce picoCTF flags from the challenge description alone (Section~\ref{sec:eval-rq3-mem}). The behavioral profiles match this divergence: picoCTF play is one-shot and static, whereas NYU-CTF is an iterative, dynamic grind (Appendix~\ref{app:beh-datasets})
\hadjer{done}\gang{Need to back up this statement about shortcuts, adding a forward pointer to your flag recall section.}

\noindent \textbf{Can configurations rescue failures.} Having diagnosed the failures, we
test whether the persistence, budget, and planner-round knobs recover them. For
each knob we report the \emph{rescue rate}: the fraction of a baseline's
same-strategy failures that the modified configuration now solves
(Table~\ref{tab:rescue}). The interventions are well-matched to the diagnosis: a
no-give-up prompt for the give-up-prone codex, a larger budget for the
budget-bound Claude models. Rescue rates range from only 10--15\% of
failures. Budget increases the budget allocation per challenge from \$3 to \$5 and rescues 0.14 of opus's vanilla failures;
persistence rescues 0.10--0.15 of codex's give-up failures. Raising the planner rounds cap from 30 to 50 helps
only vanilla gpt5 (0.11). Raising any single allowance rarely converts a failure into a solve (across budget, rounds, and give-up knobs). The unrescued runs either re-exhaust the larger allowance or shift to the next binding limit: give-up→rounds for codex, rounds→budget for gpt5. This can allude to hitting a competence wall not resource caps.


\subsubsection{Failure Signatures}
\label{sec:eval-rq2-behav}
We next ask what distinguishes a successful run from a failed one, using
\reveree's Tier-3 features on the merged NYU-CTF set. We summarize by the results by computing the Cliff value
as solved minus failed, so a positive sign marks a behavior enriched in successful runs. We exclude picoCTF from this comparison: its challenges are mostly easy and source-based, so a representative run is short, and only a few atypically long runs carry enough behavioral signal to profile.

\noindent\textbf{What separates success from failure.} One behavioral contrast holds across all models with varied prompting strategies, versions, sizes, and families. At the level of individual features,
successful runs are distinguished by the \emph{productive pivot}: moving from
analysis into building a solver (Cliff's~\cite{Cliff} $\delta=+0.22$ to $+0.38$). Additionally, by the time spent scripting ($+0.09$ to $+0.34$) and the pivot rate ($+0.04$
to $+0.23$). On the other hand, failed runs are instead marked by churn/being stuck: command repetition
($\delta=-0.36$ to $-0.48$), repeated whole-program re-analysis ($-0.21$ to
$-0.26$), and time lost to dynamic execution ($-0.16$ to $-0.30$). 
The same distinction appears as concrete action sequences. The patterns consistently occurring in successful runs are all pivots into
scripting: \textsc{static}$\to$\textsc{script},
\textsc{static}$\to$\textsc{static}$\to$\textsc{script}, and
\textsc{script}$\to$\textsc{script}. Failed runs, in contrast, loop on re-reading and
triage: \textsc{static}$\to$\textsc{triage},
\textsc{static}$\to$\textsc{static}, and
\textsc{static}$\to$\textsc{static}$\to$\textsc{static}. 

 \gang{Is there a table or figure we can look at for these results? If we run out of space, put figures/tables to appendix while discussing them here.}

\insight{
How an agent fails depends on the model, not only the challenge. On hard challenges the bottleneck is comprehension; on easy ones, solved runs skip analysis resulting in shortcut solves. Raising allowances rescues few failures, pointing to a competence wall. Success and failure also have a consistent shape across models and strategies: successful runs pivot from analysis into building a solver, whereas failed runs re-analyze without advancing. Strategies that help do so by inducing this behavior: more validation, less re-analysis.}


    

\subsection{RQ3: Reasoning vs.\ Memorization}
\label{sec:eval-rq3}

Public CTF challenges and their walkthroughs are available online and thus are likely included in the training data of models tested in
Section~\ref{sec:eval-setup}. We therefore ask whether the competence measured for the models mentioned above is genuine reverse engineering or sensitivity to memorized public
benchmarks. We hold the underlying RE task constant and perturb only the features that a memorizing or shortcut-reliant model would depend on, separating two channels: \emph{flag recall} and
\emph{surface mutations}.

\subsubsection{Flag Recall}
\label{sec:eval-rq3-mem}

We first ask each model to produce a challenge's flag given only its description, {\em without providing the binary or any other files/tools}. In this case, it should be almost impossible for the model to solve the challenges without memorizing the solutions from its training data.  
We measure both {\em exact matches} of the flag and {\em partial matches} of the flag (i.e., the flag given by the model shares at least four consecutive characters with the real flag). 
PicoCTF results indicate memorization of the models. For example, Opus 4.7 reproduces 7.4\% of picoCTF's exact, complete flags (25.9\% partial flags), without accessing the binaries at all. Gpt5 and codex reach 25.9\% and 18.5\% partial recall with no exact matches. On the NYU-CTF dataset the recall is near zero (at most 3.3\% partial and no exact matches). The result indicates that picoCTF suffers more from model memorization. 


\begin{table}[t]\centering
\small
\begin{tabular}{l l l cc}
\toprule
Pert. & Dataset & Model & Solve (o$\to$p) & Stage (o$\to$p) \\
\midrule
\multirow{6}{*}{Name}
 & \multirow{3}{*}{picoCTF} & opus 4.7     & $1.00 \to 0.96$ & $0.59 \to 0.54$ \\
 &                          & codex        & $\textbf{1.00} \to \textbf{0.95}$ & $0.72 \to 0.64$ \\
 &                          & gpt5         & $\textbf{1.00} \to \textbf{0.95}$ & $\textbf{0.71} \to \textbf{0.59}$ \\
\cmidrule(l){2-5}
 & \multirow{3}{*}{NYU-CTF} & opus 4.7     & $1.00 \to 0.92$ & $0.69 \to 0.63$ \\
 &                          & codex        & $1.00 \to 0.81$ & $0.75 \to 0.75$ \\
 &                          & gpt5         & $\textbf{1.00} \to \textbf{0.70}$ & $\textbf{0.70} \to \textbf{0.59}$ \\
\midrule
\multirow{6}{*}{Desc.}
 & \multirow{3}{*}{picoCTF} & opus 4.7     & $1.00 \to 1.00$ & $0.59 \to 0.56$ \\
 &                          & codex        & $\textbf{1.00} \to \textbf{0.95}$ & $0.72 \to 0.68$ \\
 &                          & gpt5         & $\textbf{1.00} \to \textbf{0.95}$ & $\textbf{0.71} \to \textbf{0.66}$ \\
\cmidrule(l){2-5}
 & \multirow{3}{*}{NYU-CTF} & opus 4.7     & $1.00 \to 0.92$ & $0.69 \to 0.68$ \\
 &                          & codex        & $1.00 \to 0.87$ & $\textbf{0.75} \to \textbf{0.60}$ \\
 &                          & gpt5         & $\textbf{1.00} \to \textbf{0.75}$ & $0.70 \to 0.61$ \\
\bottomrule
\end{tabular}
\caption{Effect of perturbing the name and description split by
dataset: solve rate and stage coverage, original~$\to$~perturbed, each model scored on its
own solved baseline set. A decline indicates reliance on the perturbed cue.
\hadjer{\textbf{[DONE]}}
\gang{Can we bold the biggest declines per setting?} 
\gang{If you } 
}
\label{tab:rq3}
\end{table}

\subsubsection{Surface Mutations}
\label{sec:eval-rq3-mut}
We measure the solve-rate drop from surface perturbation alone, scoring each model on
the challenges it solved at baseline. For each model we choose the samples that were solved by the baseline (vanilla), apply the perturbations to these samples and then rerun the experiments on the perturbed samples without changing anything else. This allows us to measure/isolate the solve-rate drop from surface perturbation alone.
These perturbations, including renaming the challenges and paraphrasing the question description, are done manually and do not change the semantics of the challenge itself.  

Table~\ref{tab:rq3} reports solve rate and stage coverage, original$\to$perturbed, split by dataset. On
picoCTF, renaming and paraphrasing change solve rate by at most $5$ points for any
model. On NYU-CTF, solve rate drops $7.7$--$30.0$ points under the same two conditions:
opus 4.7 falls $7.7$ points under both; codex falls $18.8$ points (renamed) and $12.5$ points (paraphrased); gpt5 falls $30.0$ points (renamed) and $25.0$ points (paraphrased). Stage coverage shows the same split: near-zero change on picoCTF, larger
declines on NYU-CTF except codex's renamed-condition stage coverage.
This shows that perturbations have an effect on the model's ability to solve challenges, but the majority of the solves were still maintained. The result indicates that while the memorization effect exists, there is still a level of genuine reasoning to solve these RE tasks.

\gang{I think we want to adjust/tune down the claims. We should at least acknowledge that the memorization effect exists. At the same time, we can say there is a genuine effort to reason and solve the RE tasks despite the memorization effect. I slightly edited the last sentence and you can do another pass. }

\insight{Memorization is present and benchmark-dependent: picoCTF flags are recalled from descriptions alone; NYU-CTF shows minimal recall. Agents also lean on surface cues, but most solves survive perturbation. This means genuine analysis coexists with memorization, rather than being replaced by it. To use easy, public benchmarks, contamination-controlled evaluation is required.}

\section{Related Work}
\label{sec:related}

\noindent\textbf{LLM agents for CTF and offensive security.}
Autonomous CTF agents are evaluated almost entirely by \emph{solve rate}:
NYU-CTF~\cite{nyuctf}, Cybench~\cite{cybench}, picoCTF~\cite{yang2023intercode},
EnIGMA~\cite{enigma}, D-CIPHER~\cite{dcipher}, and plain-agent baselines~\cite{turtayev} all rank systems by the fraction of challenges whose flag is captured. EnIGMA equips a single agent with interactive terminal tools ~\cite{enigma}; PentestGPT traces failure to context loss over long runs~\cite{pentestgpt}; D-CIPHER adds a planner--executor multi-agent harness~\cite{dcipher}. 
We contribute no agent; instead, we propose a method that evaluates agents' runs based on an RE-specific stage schema end-to-end. We hold the D-CIPHER harness fixed and vary the base model, prompting strategy, and configurations. 


\noindent\textbf{Language models for reverse engineering.}
A parallel line targets individual RE primitives rather than end-to-end challenge solving. These includes tasks such as neural and LLM decompilation (LLM4Decompile~\cite{llm4decompile}, SLaDe~\cite{slade}, Idioms~\cite{idioms}, DeGPT~\cite{degpt}, Nova~\cite{nova}), recovery of names, types, and structures from stripped binaries (DIRTY~\cite{dirty}, ReSym~\cite{resym}, GenNm~\cite{gennm}, SymLM~\cite{symlm}, XFL~\cite{xfl}), function-name inference~\cite{symgen}, GNN-based type inference TYGR~\cite{tygr}, disassembly and instruction representation (XDA~\cite{xda}, Loadstar~\cite{Benkraouda2025YouCJ}, D-ARM~\cite{darm}, PalmTree~\cite{li2021palmtree}, jTrans~\cite{wang2022jtrans}), summarization and deobfuscation (BinT5~\cite{bint5}, Bin2Summary~\cite{song2024bin2summary}, Beste et al.~\cite{deobfuscation}), and benchmarks of LLM binary understanding~\cite{howfar}. We instead evaluate agents that must orchestrate a full toolchain across a challenge end-to-end. If these tools are incorporated into the toolchain the agent has access to, they become one stage in a longer trajectory we score.


\noindent\textbf{Evaluating agents beyond solve rate.}
CTFJudge scores agent trajectories with an LLM judge against expert write-ups across all CTF categories, introducing a Competency Index for partial correctness~\cite{ctfjudge}. This is done using generic competency dimensions in a single whole-trajectory LLM pass. \reveree, on the other hand, focuses on RE challenges and evaluates agents' runs based on an RE-specific stage schema. Additionally, it confines the LLM judge to the comprehension stages of a reverse-engineering workflow. It requires the LLM judge to cite evidence for every award, blinds the judge to the outcome,  and is validated against a human rater. Mechanical stages are evaluated deterministically. Another differentiating factor is the behavioral
profiling provided by \reveree. The broader code-agent literature supplies the precedent for judging trajectories at scale (SWE-agent~\cite{sweagent}; Cemri et al.'s MAST taxonomy of multi-agent failures~\cite{mast}); we adopt standard cautions on LLM-as-judge reliability~\cite{zheng2023judging,wang2024fairevaluators}.

\noindent\textbf{Failure analysis.} Nishizaka et al. qualitatively attribute failures on 24 ``crackmes'' across three commercial agents to four weaknesses: training bias, over-trust in observations, context limitation, and plan persistence. Their analysis uses a three-stage (Observe--Comprehend--Plan) loop and qualitative labeling. Two evaluators hand-tag which protection exposes which weakness~\cite{nishizaka}.
\reveree differs in granularity, scale, and output. \reveree computes partial solves and generates a behavioral profile \emph{(output)} based on an eight-stage RE schema \emph{(granularity)} across 88 challenges and nine models and four prompting strategies \emph{(scale)}. The nine-model, four-strategy scale is what licenses an invariance claim that \cite{nishizaka}'s three-agent study cannot make, and the behavioral tier reports the run-level \emph{dynamics} of failure
not only its cognitive cause. 

\noindent\textbf{RE reasoning and memorization.}
Because CTF write-ups leak into pre-training, solve rate conflates capability with recall. A common mitigation perturbs benchmark items with
semantics-preserving transformations and reads the drop as a contamination signal. Prior work suggests generic LLM-benchmark mutation methods that do not apply directly to RE
~\cite{varbench, riddell2024quantifying},
and they expose a fidelity--resistance tradeoff in which heavy rewrites distort the task~\cite{fidelity}.
An RE-specific precedent EnIGMA~\cite{enigma}, does not perturb the challenge at all: it flags leakage \emph{post hoc} from agent runs in two ways: 1) by detecting runs that submit the flag in a single step or without analysis and 2) by estimating contamination from a separate held-out set of post-cutoff challenges. The held-out comparison spans \emph{different} challenges which can confound contamination with difficulty and the
trajectory test registers only the most blatant recall. We instead adapt the perturbation idea to RE, mutating at the \emph{challenge} level (renaming, paraphrasing), since the artifact an agent consumes is the compiled
binary. This yields a difficulty-matched original-versus-perturbed pair.

\section{Discussion}
\label{sec:discussion}

\subsection{What Generalizes Beyond These Models}
The rankings, solve rates, and costs above are a snapshot that the next model
generation will overwrite; what survives is \reveree and the structural
regularities it exposes. As an evaluation framework, it asks where an agent fails, how it
behaves, and whether success reflects analysis or recall, which will be 
applicable to 
future models.
Three findings are claims about the task, not one model's
score, and should persist: a single solve rate conflates stage progress with flag
capture (Section~\ref{sec:eval-rq1-models}); the binding constraint is localizable and 
presently competence, not resources. Specifically, failures concentrate at comprehension stages, while budget, persistence, and round-limit interventions recover
only $10$--$15\%$ of them (Section~\ref{sec:eval-rq2-fail}); and productive and
unproductive trajectories have a characteristic shape: validate-then-script
versus re-read-and-loop. These patterns are replicated across all nine models (Section~\ref{sec:eval-rq2-behav}). The evaluation lessons strengthen with time: an easy benchmark hides genuine prompting effects (Section~\ref{sec:eval-rq1-strat}), and as write-ups leak ever further into pre-training (Section~\ref{sec:eval-rq3}), a held-out split and a surface-mutation protocol become \emph{more} necessary, not less.
Re-run on the next generation, the framework reports not a new number but \emph{what changed}: where the bottleneck moved, whether the signature holds, and whether scaffolding finally pays off.

\subsection{Security-Specific Prompting Strategies}
\label{sec:disc-prompting}

The four strategies we evaluate were built for other domains, where they deliver
large gains: chain-of-thought lifts GSM8K accuracy by roughly forty
points~\cite{cot}, and ReAct and Reflexion improve embodied task success by
twenty to thirty-four points~\cite{react,reflexion}. On reverse engineering tasks, however, the
same strategies move solve rate by under three points on the easy picoCTF
challenges and by seven to twelve points on the harder NYU-CTF challenges, and
overlap so heavily that $72\%$ of solved cases are solved by all
four (Section~\ref{sec:eval-rq1-strat}). Prompting tuned for arithmetic or web
navigation, in short, transfers poorly to RE. In other security domains, domain-specific \emph{prompting} has been developed. For example, malware-scoped hierarchical prompting reaches $77\%$ against $49.5\%$
for a generic prompt~\cite{malparse}, and vulnerability-semantics chain-of-thought raises macro-F1 from $3\%$ to $36\%$~\cite{vsp}. Yet no RE-specific prompting strategies have been developed. A reasoning schema built for the
comprehension stages, rather than a generic strategy imported as-is, is a concrete and largely unexplored direction.

\subsection{Limitations and Future Work}
Our work has a few limitations, each pointing to a future direction. First, every solve rate is single-attempt: one trajectory per (model, strategy,
challenge), without pass@$k$ or resampling. Within a fixed budget, we prioritized
breadth (9 models, 4 prompting strategies, 88 challenges) over repeated sampling. We anchor our claims on effects stable across the corpus: the comprehension bottleneck and
the success--failure signature hold across all nine models.
Estimating within-configuration variance under pass@$k$ on a
stratified subset is a natural next step. Second, we instantiate \reveree{} on a single harness, \dcipher{}.
\reveree{} is harness-agnostic, so replicating the analysis on other harnesses (e.g., \textsc{EnIGMA}~\cite{enigma}) would be a natural next step. Finally, our corpus spans two sources that behave differently: picoCTF and NYU~CTF. This asymmetry is itself a finding.
Scaling the evaluation with larger and more diverse challenge sets (e.g., harder binaries), and more difficulty-preserving perturbations would both sharpen \reveree{}'s contamination signal.

\section{Conclusion}
\label{sec:conclusion}

This paper introduced \reveree, a framework for evaluating LLM reverse-engineering agents beyond solve rate. \reveree scores each run at three tiers, solve rate, milestone progress through an eight-stage RE schema, and a behavioral profile of the agent's actions, and pairs them with a memorization probe.
Applying it to nine frontier models under four prompting strategies on 88 picoCTF and NYU-CTF challenges, we found that the base model dominates performance and that larger, newer, or costlier models are not reliably stronger. We also found that failures concentrate at the comprehension stages and reflect a competence limit rather than a resource limit, while successful runs share a behavioral signature that failed runs lack. Finally, 
regarding memorization, models recall picoCTF flags from challenge descriptions alone whereas NYU-CTF shows near zero recall, and most solves survive surface perturbation, so analysis coexists with memorization rather than being replaced by it. The lasting contribution is the methodology: \reveree{} can be applied to future agents and its diagnostic questions grow more relevant as contamination spreads. We release \reveree to support future work.

\gang{TODOs: (1) write a paragraph to say we open source our dataset and code. (2) We must include an \url{https://anonymous.4open.science} link in this section, using the conference ID ``SEC27'' to apply the current cycle’s expiration deadline. Check any expiration times carefully: files must remain accessible for the entire evaluation period, until the shepherd approval deadline.
(3) make sure the link is 100\% anonymous, otherwise desk reject.
(4) 3-day grace period for the code/data artifacts. After 3 days past deadline, must be frozen. Update (showing on timestamp) will be desk reject. }

\bibliographystyle{plain}
\bibliography{sections/refs_final}
\clearpage

\appendix
\appendix

\section{Behavioral Profiles: Definitions and Additional Results}
\label{app:behavior}
This section provides extra details on the Tier~3 behavioral profile of Section~\ref{sec:reveree-behavior} and related results. Appendix~\ref{sec:markov-matrix} defines the components of the profile that Section~\ref{sec:reveree-behavior} introduces only by name, all derived from the coarse-state transition matrix of a run. The remaining subsections slice the profile along the three factors varied in RQ1 (Section~\ref{sec:eval-rq1}): Appendix~\ref{app:beh-models} reports per-model fingerprints, Appendix~\ref{app:beh-datasets} contrasts picoCTF and NYU-CTF play, and Appendix~\ref{app:beh-strategies} shows how the prompting strategies that help on NYU-CTF change what the agent does. None of these results alters an RQ1 conclusion; they explain \emph{how} the differences reported there arise, and the last of them supplies the mechanism behind the success-versus-failure signature of Section~\ref{sec:eval-rq2-behav}.
\subsection{Metric Definitions}
\label{sec:markov-matrix}
The \emph{static self-transition}, $n_{\textsc{Static},\textsc{Static}} / \sum_k
n_{\textsc{Static},k}$, measures how often the agent stays in static analysis from one
step to the next, i.e., unproductive ``grinding.'' The \emph{re-analysis count} is the
number of times the agent re-runs whole-program analysis from scratch instead of reusing
what an earlier agent invocation already produced.
also reported as a rate normalized by the number of static actions. \emph{Command repetition} is the fraction of issued shell commands that re-issue a previously seen, normalized command signature. The \emph{pivot rate} is the fraction of transitions that change state, and the \emph{productive pivot}, $n_{\textsc{Static},\textsc{Script}} / \sum_k
n_{\textsc{Static},k}$, is the empirical probability of moving into scripting conditional on leaving static analysis. 
The remaining components track how far the run moves beyond static reading: the static-to-dynamic action ratio and the position of the first dynamic action capture when and how much it turns from reading code to executing it, and two binary indicators record whether the run ever reached scripting (building a solver) or validation.

\subsection{Model Behavior} 
\label{app:beh-models}
\begin{figure*}[h]
\centering
\includegraphics[width=0.85\linewidth]{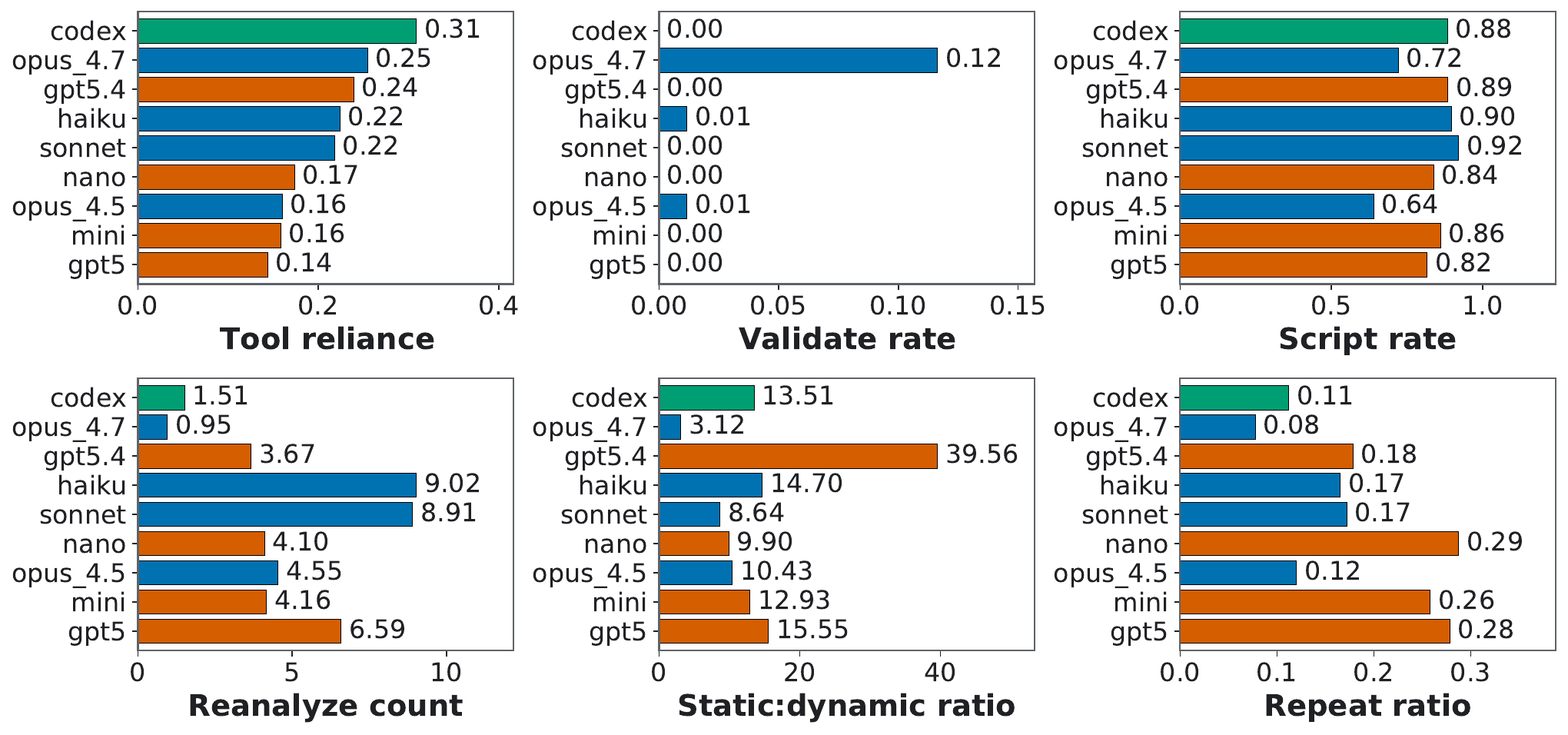}
\caption{Behavioral fingerprints per model (vanilla). Panels report Tier-3 profile features averaged over challenges.}
\label{fig:b3}
\end{figure*}

Models also differ in \emph{how} they work, not only in how often they succeed (Fig.~\ref{fig:b3}). For example, models differ on how much tool reliance, the fraction of actions issued through decompilation, disassembly, or scripting rather than raw shell, they exhibit. Codex has a tool reliance value of 0.309, the most tool-reliant, while gpt5 at 0.144 is the most bash-centric. Additionally, we see differing profiles with respect to the coarse states reached by each model. Opus is the only model that reaches \textsc{validate} with significant frequency, i.e.\ that checks a candidate answer before submitting it. These fingerprints are stable across challenges.

\subsection{Dataset Behavior}
\label{app:beh-datasets}
\begin{figure}[t]
\centering
\includegraphics[width=\linewidth]{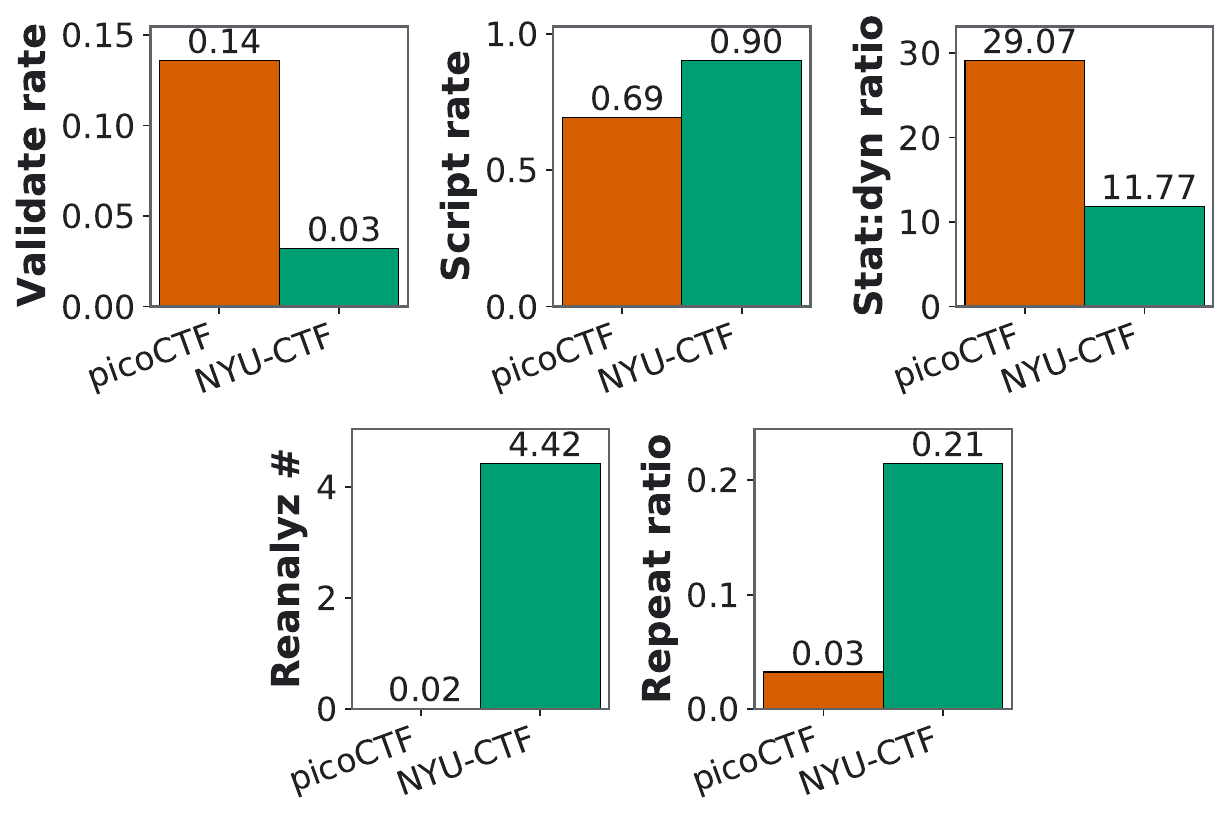}
\caption{Behavioral profile per dataset. Each panel reports one Tier-3 profile feature, with one bar for picoCTF and one for NYU-CTF.}
\label{fig:dsbeh}
\end{figure}

The behavioral profiles match this divergence (Fig.~\ref{fig:dsbeh}). PicoCTF play is one-shot and static: a static-to-dynamic action ratio of about 29, almost no re-analysis (0.02 passes per run), and the highest validation rate (0.14). NYU-CTF is an iterative, dynamic grind: a static-to-dynamic ratio of about 12, 4.42 re-analysis passes per run, and heavy scripting (0.90). At the model level, re-analysis count is the sharpest discriminator on NYU-CTF. gpt5 re-reads artifacts about 9.55 times per run against opus 4.7's 1.39 (codex 2.20, gpt5.4 5.32) and codex again shows the deepest NYU-CTF progress depth (0.414) at the lowest solve rate (0.276).

\subsection{Prompting Strategy Behavior}
\label{app:beh-strategies}
\begin{figure*}[t]
\centering
\includegraphics[width=\linewidth]{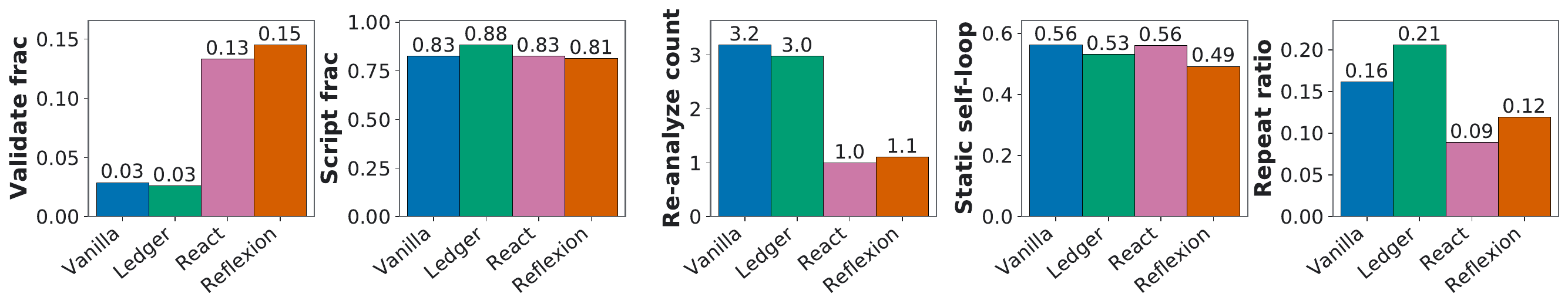}
\caption{Behavioral profile by prompting strategy. Each panel reports one Tier-3 profile feature averaged over the four multi-strategy models.}
\label{fig:bxs}
\end{figure*}

The strategies that help, help by changing what the agent does (Fig.~\ref{fig:bxs}). ReAct and Reflexion reach \textsc{validate} far more often than vanilla or ledger (0.13 and 0.15 versus 0.03) and re-analyze the binary about three times less often (1.0 and 1.1 passes per run versus 3.2 and 3.0). In other words, the strategies that win do so by checking answers and avoiding redundant re-analyzing, a connection we make precise at the level of success and failure in Section~\ref{sec:eval-rq2-behav}.

\section{Judge Validation Sample}
\label{app:validation-sample}
Table~\ref{tab:validation-sample} lists the 15 runs used in the human agreement study of Section~\ref{sec:judge-validation}, identified by dataset, outcome, agent, and challenge. The sample was drawn at random from the judged corpus subject to covering both benchmarks, both outcomes, and all three model families; the annotation procedure and results are reported in Section~\ref{sec:judge-validation}.

\begin{table}[h]
  \centering
  \footnotesize
  \begin{tabular}{rllll}
    \toprule
    \# & Dataset & Outcome & Agent & Challenge \\
    \midrule
    1  & NYU-CTF & unsolved & opus 4.7  & 2020q-rev-ezbreezy \\
    2  & NYU-CTF & unsolved & codex     & 2020q-rev-ezbreezy \\
    3  & NYU-CTF & unsolved & gpt5      & 2019f-rev-wrangly \\
    4  & NYU-CTF & unsolved & nano      & 2019f-rev-steady\_counting \\
    5  & NYU-CTF & solved   & opus 4.7  & 2023q-rev-whataxor \\
    6  & NYU-CTF & solved   & sonnet    & 2023q-rev-rebug\_2 \\
    7  & NYU-CTF & solved   & codex     & 2023q-rev-rebug\_2 \\
    8  & NYU-CTF & solved   & codex     & 2021q-rev-checker \\
    9  & picoCTF & solved   & opus 4.7  & 2025q-rev-10 \\
    10 & picoCTF & solved   & gpt5      & 2025q-rev-10 \\
    11 & picoCTF & solved   & sonnet    & 2025q-rev-52 \\
    12 & picoCTF & solved   & gpt5.4    & 2025q-rev-44 \\
    13 & picoCTF & solved   & gpt5.4    & 2025q-rev-46 \\
    14 & picoCTF & unsolved & codex     & 2025q-rev-15 \\
    15 & picoCTF & unsolved & gpt5      & 2025q-rev-43 \\
    \bottomrule
  \end{tabular}
  \caption{Runs in the human agreement sample: 15 runs over 12
    unique challenges, chosen at random subject to covering both
    benchmarks, both outcomes, and all two agent families. 
    \gang{Can be moved to Appendix if we need space.}
    }
  \label{tab:validation-sample}
\end{table}

\end{document}